\documentclass[showpacs,preprintnumbers,amsmath,amssymb, 
eqsecnum,
twocolumn, tightenlines,
]{revtex4}

\usepackage{bm}

\usepackage{amsmath}

\usepackage{dcolumn}

   \usepackage[dvips]{graphicx}
\begin{document}

    \bibliographystyle{apsrev}

    \title {Coulomb problem for vector bosons}

    \author{M.Yu.Kuchiev} 
    \email[Email:]{kuchiev@phys.unsw.edu.au}
    \author{V.V.Flambaum}
    \email[Email:]{flambaum@phys.unsw.edu.au} 
    \affiliation{School of Physics, University of New South Wales,
      Sydney 2052, Australia} 
    \affiliation{Physics Division, Argonne
      National Laboratory, Argonne, Illinois 60439-4843, USA}

    \date{\today}

    \begin{abstract} 
      The Coulomb problem for vector bosons $W^\pm$ incorporates a
      well known difficulty; the charge of the boson localized in a
      close vicinity of the attractive Coulomb center proves be
      infinite.  This fact contradicts the renormalizability of the
      Standard Model, which presumes that at small distances all
      physical quantities are well defined. The paradox is shown to be
      resolved by the QED vacuum polarization, which brings in a
      strong effective repulsion that eradicates the infinite charge
      of the boson on the Coulomb center. This property allows to
      define the Coulomb problem for vector bosons properly, making it
      consistent with the Standard Model.
    \end{abstract}


    \maketitle

    \section{introduction}
    \label{intro}
    
    Consider a charged vector boson, which propagates in the Coulomb
    field created by a heavy point-like charge $Z$ assuming that the
    boson is massive, its mass being produced via the Higgs mechanism;
    the $W^\pm$-bosons give an example. We study relativistic effects
    in this Coulomb problem. A situation where they can be important
    arises, for example, for small primordial charged black holes
    since an impact of their Coulomb field on a $W$-boson prevails
    over the gravitational field.
          
    It has ``always'' been known that there is a difficulty in the
    Coulomb problem for vector bosons.  Soon after Proca formulated
    theory for vector particles \cite{proca_1936} it became clear that
    it produces inadequate results for the Coulomb problem
    \cite{massey-corben_1939,oppenheimer-snyder-serber_1940,tamm_1940-1-2}.
    This fact inspired Corben and Schwinger
    \cite{corben-schwinger_1940} to modify the Proca theory, tuning
    the Lagrangian and equations of motion in such a way as to force
    the hyromagnetic ratio of the vector boson to acquire a favorable
    value $g=2$.  Later on the formalism of
    \cite{corben-schwinger_1940} was found to have a connection with
    the non-Abelian gauge theory \cite{schwinger_1964}, which makes it
    relevant for the present day studies.  A role of the identity
    $g=2$ was thoroughly discussed in literature, see e. g.
    Ref.\cite{cheng_wu_1972,huang_1992}.
   
    Ref.\cite{corben-schwinger_1940} found a realistic discrete energy
    spectrum for the Coulomb problem for vector bosons. However, it
    discovered also a fundamental flaw in the problem. For two series of
    quantum states the charge of the vector boson located on the
    Coulomb center turns infinite, which indicates the fall of the
    boson on the center.  One of these series has the total angular
    momentum zero, $j=0$, another one has $j=1$ (being further
    specified by a label ``$\gamma-3/2$'', see Section \ref{j>0}).
    This effect takes place for arbitrary small value of the Coulomb
    charge $Z$, which is physically unacceptable.  Moreover, it takes
    place at small distances, while the renormalizability of the
    Standard Model Ref.\cite{thooft-veltman-1972} guarantees that
    there should be no problems of this type. All this indicates that
    the Coulomb problem is poorly defined. Moreover, there exists a
    contradiction; the Coulomb problem derived from the Standard Model
    produces results, which challenge the Model itself.
    
    This difficulty was inspirational for several lines of research.
    Early efforts are summarized in
    Ref.\cite{vijayalakshmi-seetharaman-mathews_1979}.  More recent
    Refs.  \cite{pomeransky-khriplovich_1998,pomeransky-se'nkov_1999,%
      pomeransky-sen'kov-khriplovich_2000} suggested a new, refined
    modification of the formalism for vector bosons.
    Ref.\cite{silenko_2004} claimed that it complies with results of
    Corben and Schwinger. Some authors considered other forms of the
    equation governing vector bosons
    \cite{fushchych-nikitin-susloparow_1985,fushchych-nikitin_1994,%
      sergheyev_1997}, which produce more acceptable results for the
    Coulomb problem, but this advantage is partially undermined by the
    fact that it does not step from a renormalizable theory.
    
    However, in spite of a progress made over the years, there still
    exists a contradiction between the difficulty in the Coulomb
    problem for vector bosons and the renormalizability of the
    Standard Model. We find a clear way to resolve this contradiction,
    formulating the Coulomb problem for vector particles properly,
    within the frames of the Standard Model. Our main observation is
    that the polarization of the QED vacuum has a profound impact in
    the problem forcing the density of charge of a vector boson to
    decrease at the origin, thus making the Coulomb problem stable,
    well defined. This decrease has an exponential character for the
    $j=0$ state. For the $j=1$,``$\gamma-3/2$'' state the suppression is
    of a power-type. In both these states the suppression eradicates
    the difficulty of the Coulomb problem.
    
    From the first glance this result looks surprising. Presumably,
    the vacuum polarization is meant to make the attractive Coulomb
    field only stronger, which should result in an increase of the
    charge density at the origin.  In addition to this, the vacuum
    polarization for spinor and scalar particles in the Coulomb field
    is known to produce only small, perturbative effects.  In
    contrast, we claim a strong {\it reduction} of the charge density
    for the vector particle. To grasp a physical mechanism involved it
    is necessary to notice that the equation of motion for vector
    particles incorporates a particular term, which explicitly depends
    on the external current and has no counterparts for scalars and
    spinors, see the last term in Eq.(\ref{form}).  Precisely this
    term brings in a strong effective repulsion, which stems from the
    vacuum polarization and makes the Coulomb problem stable, well
    defined.
    
    The renormalizability of the Standard Model means that if all
    essential processes are taken care of, then the infinite charge of
    a vector boson located at the Coulomb center is eliminated.  It is
    known that the amplitude of the photon exchange between leptons
    or/and quarks at high transferred momenta should be considered
    alongside exchange by the Higgs and $Z$-bosons. From this
    perspective the catastrophic behavior of the charge density of a
    vector boson at small distance, i. e. at large transferred
    momenta, in the Coulomb problem could have been considered as an
    indication that the Coulomb problem for vector bosons should
    include the processes related to the Higgs and $Z$-bosons exchange
    from the very beginning. In contrast to this widely spread
    presumption we find a way to formulate the Coulomb problem for
    vector bosons entirely in terms of the $W$ and electromagnetic
    fields, as a pure QED problem.
    
    A complete Standard Model calculation, where all possible
    processes are accounted for accurately, would require specific
    information on the nature of a heavy particle that creates the
    Coulomb field, i. e.  on all its quantum numbers related to the
    Standard Model. This information is not necessarily feasible. A
    simple example give primordial black holes; it is not easy to
    assert with certainty whether they have, or have not the weak
    charge, and what are their other quantum numbers in the Standard
    Model. Same questions arise in relation to other possible
    candidates for the heavy Coulomb center. As a result, a
    presumption that the exchange of the Higgs and $Z$-boson should
    play a basic role in the Coulomb problem leads to complications.
    It is fortunate therefore that the detailed information on
    properties of the heavy particle proves be redundant, that the
    Coulomb problem can be properly defined using the only physical
    parameter of a heavy particle, its electric charge.
    
    This point of view, which is advocated in the present work keeps
    the Coulomb problem simple and transparent.  On the other hand,
    it also allows one to include all other processes, which are left
    outside the scope of the Coulomb problem, by means of perturbation
    theory. Our preliminary calculations indicate that the exchange by
    the Higgs and $Z$-bosons, as well as possible processes with
    lepton or quark exchange, give only small corrections. The reason
    stems from the fact that the found wave functions for vector
    bosons are suppressed at small distances.  Consequently, the
    small-distance processes with the exchange by Higgs and $Z$-bosons
    are also suppressed (the exchange by a lepton or quark contains
    the vanishing at the Coulomb center fields, which describe the
    $W$-boson).

    In section \ref{vector} the Corben-Schwinger formalism for charged
    vector bosons is derived directly from the Standard Model. The
    pure Coulomb problem is discussed in Sections
    \ref{Nonrel}-\ref{catastroph} and several Appendixes.  This
    analyses follows Ref.\cite{corben-schwinger_1940}, but some
    important details, including the non-relativistic limit (Section
    \ref{Nonrel}) and the eigenvalue problem for $j=0$ states (Section
    \ref{matrix}) are discussed in more detail.  Sections
    \ref{vacuum_polarization},\ref{numericals} present the main result
    of the paper. They show that the QED vacuum polarization plays a
    defining role in the problem, as was first noticed in our previous
    work \cite{kuchiev-flambaum_2005}.  The units $\hbar=c=1$,
    $e^2=4\pi\alpha$ where $e<0$, are used below.

    \section{$W$-mesons in electromagnetic field}

    \label{vector}

    \subsection{$W$-bosons in Standard Model}
    \label{standard}
    
    Consider boson fields in the electroweak part of the Lagrangian of
    the Standard Model, see e.g. Ref.\cite{weinberg_2001},
   \begin{eqnarray}
      \label{gauge}
      {\mathcal L}= -\frac{1}{4}\,
      \left(\partial_\mu \boldsymbol{A}_\nu-\partial_\nu \boldsymbol{A}_\mu  +
        g \,\boldsymbol{A}_\mu \times \boldsymbol{A}_\nu\right)^2,
      \\ \nonumber
            -\frac{1}{4}\,
      \left(\partial_\mu { B}_\nu-\partial_\nu {B}_\mu  \right)^2+
      \frac{1}{2}\,D_\mu\Phi^+ D^\mu \Phi~.
    \end{eqnarray}
    Here $\boldsymbol{A}_\mu$ and $B_\mu$ are the triplet of $SU(2)$
    and the $U(1)$ gauge potentials respectively (abridged notation is
    used here).  The covariant derivative $D_\mu\Phi$ takes into
    account that the Higgs field $\Phi$ has a hypercharge $Y=2$, which
    describes its interaction with the $U(1)$ field, and is
    transformed as a doublet under the $SU(2)$ gauge transformations.
    Taking the unitary gauge one can present it via one real component
    \begin{eqnarray}
      \label{vacuum}
      \Phi =\left( \begin{array} {c} 0 \\ \phi
      \end{array} \right)~,\quad \phi=\phi^+~.  
    \end{eqnarray}
    Assuming that the scalar field develops the vacuum expectation
    value $\phi=\phi_0$ and the Higgs mechanism takes place, one finds
    that the gauge field can be presented as a new $U(1)$ field
    $A_\mu$, and a triplet of massive fields $W^\pm_\mu, \,Z_\mu$
    \begin{eqnarray}
      \label{Amu}
      A_\mu &=&-\sin \theta \,A_\mu^3+\cos\theta \,B_\mu~, 
      \\ \label{Zmu}
      Z_\mu &=& ~~\cos \theta \,A_\mu^3+\sin\theta \,B_\mu~,
      \\ \label{Wmu}
      W_\mu &=&~~\left(A_\mu^1-iA_\mu^2 \right)/\sqrt 2~,
    \end{eqnarray}
    Here $W_\mu\equiv W_\mu^-$ represents the $W$-boson with charge
    $e=-|e|$, and $\theta$ is the Weinberg angle.  
    
    Expanding the Lagrangian Eq.(\ref{gauge}) in the vicinity of
    $\phi=\phi_0$ and retaining only bilinear in the fields
    $W_\mu,W_\mu^+$ terms, including their interaction with the
    electromagnetic field, one derives an effective Lagrangian
    \begin{eqnarray}
      \nonumber
      {\mathcal L}^W &=& -\frac{1}{2}\left(\nabla_\mu W_\nu -
        \nabla_\nu W_\mu\right)^+ \left(\nabla^\mu W^\nu-
        \nabla^\nu W^\mu\right)  
      \\ \label{W}      
        &&+ i e \, F^{\mu\nu} \,W^+_\mu W_\nu + 
        m^2 \,W_\mu^+ W^\mu~,
    \end{eqnarray}
    which describes the propagation of $W$-bosons in an external
    electromagnetic field. Here $m$ is the mass of $W$.  The external
    field is accounted for in Eq.(\ref{W}) in the derivative
    $\nabla_\mu=\partial_\mu +i e A_\mu$ and by the term with the
    field $F^{\mu\nu}=\partial^\mu A^\nu-\partial^\nu A^\mu$. The
    first and the last terms in Eq.(\ref{W}) are present in the Proca
    formalism \cite{proca_1936}, while the second one was introduced
    by Corben and Schwinger \cite{corben-schwinger_1940}.
    
    From Eq.(\ref{W}) one derives the classical Lagrange equation of
    motion for vector bosons
    \begin{eqnarray}
      \label{wave}
      \left( \nabla^2+m^2\right) W^\mu
      + 2 i e \,F^{\mu\nu}\,W_\nu-  
      \nabla^\mu \nabla^\nu \,W_\nu =0~.
    \end{eqnarray}
    Here an identity $[\nabla_\mu,\nabla_\nu]=ieF_{\mu\nu}$ was used.
    Taking a covariant derivative in Eq.(\ref{wave}) one finds
    \begin{eqnarray}
      \label{Lgauge}
      m^2\nabla_\mu W^\mu+ie\,j_\mu W^\mu=0~,
    \end{eqnarray}
    where 
    \begin{eqnarray}
      \label{j}
      j^\mu=\partial_\nu F^{\nu\mu}~.
    \end{eqnarray}
    is the external current, which creates the external field
    $F^{\nu\mu}$.  Evaluating $\nabla_\mu W^\mu$ from
    Eq.(\ref{Lgauge}) and substituting the result back into
    Eq.(\ref{wave}) one rewrites the latter one in a more transparent
    form
    \begin{eqnarray}
      \label{form}
      \left( \nabla^2+m^2\right) W^\mu
      + 2 i e F^{\mu\nu}W_\nu 
      +\frac{ie}{m^2} \nabla^\mu (j_\nu W^\nu) =0.~~\quad
    \end{eqnarray}
    This equation of motion for vector bosons was suggested in
    Ref.\cite{corben-schwinger_1940}. The coefficient 2 in front of
    the second term ensures that the g-factor of the boson takes the
    value $g=2$, see Eq.(\ref{moment}) below.
    
    The derivation outlined shows that Eq.(\ref{form}) represents the
    classical equation of motion for $W$-bosons in the external
    electromagnetic field, which is valid within the frames of the
    Standard Model.  This equation has similarities with the
    Klein-Gordon and Dirac equations (if the latter one is written as
    the second-order differential equation), but there is also an
    important distinction. It is produced by the last term in
    Eq.(\ref{form}), which explicitly contains the external current;
    there is no similar terms for scalars and spinors. We will see how
    important this term is, when we discuss the vacuum polarization.
    
    We will use below a current of vector bosons
    $j_{\phantom{\,}\mu}^{W}$, which can be obtained by considering a
    variation of the Lagrangian Eq.(\ref{W}) under variation of
    $A_\mu$, which yields
    \begin{eqnarray}
      \label{jjj}
      j_{\phantom{\,}\mu}^{W}&=&j_{\mu}^{(1)}+j_{\mu}^{(2)}+j_{\mu}^{(3)}~,
      \\ \label{j1}
      j_{\mu}^{(1)} &=& -ie\,( \,W^+_\nu \nabla_\mu W^\nu -
      \nabla_\mu W_\nu^+W^\nu\,)
      \\ \label{j2}
      j_{\mu}^{(2)} &=& -ie\,( \,\nabla_\nu W_\mu^+W^\nu -
      W_\nu^+ \nabla^\nu W_\mu  \,)
      \\ \label{j3}
      j_{\mu}^{(3)} &=& -ie \,\partial^\nu(\, W_\mu^+ W_\nu -
      W_\nu^+W_\mu  \,)~.
    \end{eqnarray}
    Differentiating in Eq.(\ref{j3}) term by term and taking into
    account Eq.(\ref{Lgauge}) one verifies that
    \begin{eqnarray}
      \label{easy}
      j_{\mu}^{(3)} &=& j_{\mu}^{(2)} -ie( \,W_\mu^+\nabla_\nu W^\nu -
      \nabla^\nu W_\nu^+  W_\mu  \,)
      \\ \nonumber
     &=& j_{\mu}^{(2)}
     -\frac{e^2}{m^2}(\, W_\mu^{+} W_\nu+W_\nu^{+}W_\mu\,)\,j^\nu~.
    \end{eqnarray}
    Using this result, the current Eq.(\ref{jjj}) can be written in a
    compact form
    \begin{eqnarray}
      \nonumber 
      j_{\phantom{\,}\mu}^{W}&=&-ie\Big( \,W^+_\nu \nabla_\mu W^\nu + 
      2 \nabla_\nu W^+_\mu  W^\nu -c.c. \,\Big)
      \\       
      \label{jw}
     &&-\frac{e^2}{m^2}(\, W_\mu^{+} W_\nu+W_\nu^{+}W_\mu\,)\,j^\nu~.
    \end{eqnarray}
    Here $c.c.$ refers to two complex conjugated terms.
      
    \subsection{Static electric field}
    \label{static_electric_field}
    
    Consider a static electric field described by the electric
    potential $A_0=A_0(\boldsymbol{r})$ and charge density
    $\rho=\rho(\boldsymbol{r})=-\Delta A_0$. For a stationary state of
    the $W$-boson one can presume that
    \begin{eqnarray}
      \label{sta}
      \nabla_0^2
    W_\mu= -(\varepsilon-U)^2W_\mu~, 
    \end{eqnarray}
    where $\varepsilon$ is the energy of the stationary state, and
    $U=U(\boldsymbol{r})= eA_0$ is the potential energy of the
    $W$-boson in the electric field.  Eq.(\ref{Lgauge}) in this case
    gives 
    \begin{eqnarray}
      \label{simp}
      {\mathsf w}
      = ( \varepsilon-U-\Upsilon )^{-1} \boldsymbol{\nabla \cdot W} .\quad
    \end{eqnarray}
    The four-vector $W^\mu=(W_0,\boldsymbol{W})$ is presented here via
    the three-vector $\boldsymbol{W}$ and the modifies
    zeroth-component ${\mathsf w}= i W_0$.  In order to simplify notation we
    introduce also a very important for us quantity
    $\Upsilon=\Upsilon(\boldsymbol{r})$,
    \begin{eqnarray}
      \label{Ups}
      \Upsilon=\frac{e\rho}{\,m^2}=-\frac{\Delta U}{m^2}~.
    \end{eqnarray}
    Eqs.(\ref{Lgauge}),(\ref{sta}) show that this definition complies
    with (\ref{simp}).  The quantity $\Upsilon$ appears in the
    equations of motion alongside the initial potential $U$, see e.g.
    Eq.(\ref{simp}).  In this sense it plays a role of an effective
    potential energy, which is specific for vector bosons.  We will
    call it the $\Upsilon$-term, or $\Upsilon$-potential.  In this
    notation Eq.(\ref{form}) reads
    \begin{eqnarray}
      \label{wa}
       \big((\varepsilon-U)^2\!-m^2\big)\boldsymbol{W}=-\Delta
      \boldsymbol{W}-2\boldsymbol{\nabla} U
      {\mathsf w}-\boldsymbol{\nabla}(\Upsilon {\mathsf w}),&\quad\quad&
\\ 
       \label{0}
       \big((\varepsilon-U)^2\!-m^2\big) {\mathsf w}=-\Delta
            {\mathsf w}+2\boldsymbol{\nabla}U\boldsymbol{ \cdot
      W}\quad\quad\quad~~~ &&
      \\ \nonumber
      + (\varepsilon-U)\Upsilon {\mathsf w}.&&
    \end{eqnarray}
    A relation between ${\mathsf w}$ and $\boldsymbol{W}$ given by
    Eq.(\ref{simp}) shows that among four equations of motion
    Eqs.(\ref{wa}),(\ref{0}) only three are independent, precisely
    what one expects for massive vector particles.
    
    It will be useful to present Eq.(\ref{wa}) in a slightly
    different form, which can be derived by combining it with
    Eq.(\ref{simp}) and using an identity $\Delta\boldsymbol{W}
    =\boldsymbol{\nabla \times}(\boldsymbol{ \nabla \times
      W})-\boldsymbol{\nabla}(\boldsymbol{\nabla \cdot W})$, which
    gives
    \begin{eqnarray}
      \label{tran}
       \!\big((\varepsilon-U)^2\!-m^2\big)\boldsymbol{W}=
    \boldsymbol{\nabla \times}(\boldsymbol{
      \nabla \times W})
    \\ \nonumber
    -(\varepsilon-U)\boldsymbol{\nabla} {\mathsf w}\,
    -\boldsymbol{\nabla}U\,{\mathsf w}~.
    \end{eqnarray}
    From the expression for the current of vector bosons Eq.(\ref{jw})
    one derives the charge density
    \begin{eqnarray}
      \label{ro}
     \rho^W&=& 2e\Big( (\varepsilon-U)(\boldsymbol{W}^+
     \!\cdot\!\boldsymbol{W}+{\mathsf w}^+{\mathsf w})
     \\ \nonumber 
       &+&\boldsymbol{W}^+\!\!\cdot \!\boldsymbol{ \nabla} {\mathsf w}
       +
       \boldsymbol{W}\!\cdot \!\boldsymbol{ \nabla}\,{\mathsf w}^+ -\Upsilon \,
       {\mathsf w}^+{\mathsf w}\,\Big).
    \end{eqnarray}

    \subsection{G-factor}
    \label{g-factor}
    
    The behavior of vector bosons in the homogeneous magnetic fields
    was studied in detail, see e.g.
    \cite{vijayalakshmi-seetharaman-mathews_1979} and references
    therein.  The spectrum of this problem reads, see Section
    \ref{homo},
    \begin{eqnarray}
      \label{e2}
      \varepsilon^2=m^2+p_z^2+2|e|B\left(n+1/2+\sigma \right)~.
    \end{eqnarray}
    Here $n=0,1\dots$ specifies the Landau levels, and $\sigma=-1,0,1$
    gives a projection of spin $S=1$ of the vector boson.
    Eq.(\ref{e2}) shows that vector bosons possess the magnetic moment
    \begin{eqnarray}
      \label{moment}
      \mbox{\boldmath $\mu$} =e\boldsymbol{ S}/m~,
    \end{eqnarray}
    which means that the magnetic $g$-factor is $g=2$.

    \section{Non-relativistic limit}
    \label{Nonrel}
    
    Consider a vector boson in a static electric field with the
    potential energy $U=eA_0(\boldsymbol{ r})$.  If we presume that
    the non-relativistic approach is valid, which needs that $|U| \ll
    m$, then in the lowest order of the perturbation theory in powers
    of $U/m$ one immediately finds from Eqs.(\ref{wa}),(\ref{simp})
    \begin{eqnarray}
      \label{nonrel}
      E\, \boldsymbol{ W}  = - \frac{1}{2m}\,\Delta\,\boldsymbol{ W}
      +U\boldsymbol{ W} ~.
    \end{eqnarray}
    Here $E\simeq \varepsilon-m$ is the energy, the vector $\bf W$
    plays a role of the wave function for the vector boson, and the
    non-relativistic Hamiltonian on the right-hand side has a usual
    form for a massive charged particle.
    
    Let us find corrections to Eq.(\ref{nonrel}) induced by
    relativistic effects. The wave function of the massive vector
    particle ${\mbox {\boldmath $\Phi$}}$ is well defined in the rest
    frame. Therefore the vector $\bf W$, which describes the moving
    vector particle, inevitably deviates from the wave function $\mbox
    {\boldmath $\Phi$}$. A relation between $\bf W$ and $\mbox{
      \boldmath $\Phi$}$ is easy to articulate for the free motion,
    when it is given by the Lorentz boost, see e.g. a book \cite{LL4},
        \begin{eqnarray}
      \label{fw}
      \boldsymbol{ W} =\mbox{\boldmath $\Phi$}
      +\frac{\boldsymbol{ p}\,
        ( \boldsymbol{ p} \cdot \mbox{\boldmath $\Phi$}) }{m(m+\varepsilon)}
    \end{eqnarray}
    Generically, the potential energy brings in complications, but
    within the necessary accuracy we can neglect them, presuming also
    that $\varepsilon \simeq m$. Then Eq.(\ref{fw}) gives
    \begin{eqnarray}
      \label{fwapp}
      \boldsymbol{ W}&\simeq& \mbox{\boldmath $\Phi$}+\frac{\boldsymbol{ p}\,
        ( \boldsymbol{ p} \cdot \mbox{\boldmath $\Phi$}) }{2m^2}~,
    \end{eqnarray}
    where $\boldsymbol{ p}=-i\boldsymbol{ \nabla}$.  This relation
    plays a role similar to the Foldy-Wouthuysen transformation
    \cite{foldy-wouthuysen_50} for fermions.

    Substituting Eq.(\ref{fwapp}) in Eqs.(\ref{wa}),(\ref{simp}) and
    expanding the latter ones in powers of $U/m$ one finds the
    following Schr\"odinger-type equation for the wave function $\mbox
    {\boldmath $\Phi$}$ of the vector boson
    \begin{eqnarray}
      \label{corr}
      E\,\mbox{\boldmath $\Phi $}_i &=& H_{ij} \mbox{\boldmath $\Phi $}_j~, 
      \\ \label{hami}
      H_{ij}&=& \left( \frac{ \boldsymbol{ p}^2 }{2m}+U \right) 
      \delta_{ij} +\delta H_{ij}~,
    \\ \nonumber
    \delta H_{ij} &=& - \frac{ \boldsymbol{ p}^4 }{8m^3}\,\delta_{ij} 
    -\frac{  \boldsymbol{ F} \cdot (  \boldsymbol{ p \times S}_{ij} ) }{ 2m^2 }
      +\frac{\Delta  U}{6m^2} \,\delta_{ij}
    \\ \label{dH}
    &+&\frac{1}{6m^2}\left( 3 \frac{\partial^2 U}{\partial r_i
      \partial r_j}
    -\Delta
      U \,\delta_{ij}\right)~.
    \end{eqnarray}
    Here $i,j=1,2,3$ label components of three-vectors,
    $\boldsymbol{ S}$ is the spin, which operates on a vector
    $\boldsymbol{ V}$ according to $\boldsymbol{ S}_{ij}
    V_j=-i\epsilon_{ijk} V_k$. 
    
    The relativistic correction to the Hamiltonian $\delta H$ of
    vector particles is given in Eq.(\ref{dH}). It is instructive to
    compare this correction with the known Darwin Hamiltonian $\delta
    H_\mathrm{D}$, which accounts for relativistic effects for spinor
    particles
    \begin{eqnarray}
      \label{D}
    \delta H_D  = - \frac{ \boldsymbol{ p}^4 }{8m^3} 
    -\frac{  \boldsymbol{ F} \cdot (  \boldsymbol{ p \times s} ) }{ 2m^2 }
      +\frac{\Delta  U}{8m^2}~.
    \end{eqnarray}
    Here $\boldsymbol{ s}=\mbox{\boldmath $\sigma$ }/2 $ is the
    operator of spin for spinor particles. The three terms in the
    first line of Eq.(\ref{dH}) resemble their counterparts in
    Eq.(\ref{D}), the only distinction is the numerical coefficient in
    front of the term with $\Delta U$. We conclude that these three
    terms have conventional meaning, describing the relativistic
    correction to the kinetic energy, the spin-orbit interaction, and
    the contact correction to the potential. The coefficient in front
    of the term responsible for the spin-orbit interaction in
    Eq.(\ref{hami}) complies with the hyromagnetic ratio $g=2$ of the
    vector boson, if one presumes that the Thomas ``one-half rule'' is
    applicable for vector particles the same way as for spinors.
    
    The last, forth term in Eq.(\ref{dH}) finds no counterpart in the
    Darwin Hamiltonian. It is instructive to write a contribution of
    this term to the energy shift
    \begin{eqnarray}
      \label{dE}
      \delta E_Q&=&\frac{1}{6}\, \int \, Q_{ij} \,
      \frac{\partial^2 A_0}{\partial r_j
      \partial r_i} \,d^3 r~, 
    \\ \label{Q}
    Q_{ij}&=&\frac{e}{m^2} \left( 3
      \mbox{\boldmath $\Phi $}_i^*
      \mbox{\boldmath $\Phi $}_j-\delta_{ij}
      |  \mbox{\boldmath $\Phi $} |^2 \right)~.
    \end{eqnarray}
    Eqs.(\ref{dE}),(\ref{Q}) show that $Q_{ij}$ plays a role of the
    density of the quadrupole moment for vector bosons. We conclude
    that the last, forth term in Eq.(\ref{dH}) indicates that vector
    bosons have a quadruple moment.
    
    From the first glance the contact and the quadrupole terms in the
    Eq.(\ref{dH}) have similarity with the $\Upsilon$-term in
    Eq.(\ref{Ups}). However this resemblance is coincidental, since
    the $\Upsilon$-term does not contribute to (\ref{dH}), which takes
    into account corrections of the order of $(Z\alpha)^2$.
    Eq.(\ref{Ups}) allows to estimate the $\Upsilon$-potential as
    $\Upsilon\sim (mr_0)^{-2} U\sim (Z\alpha)^2 U$, where
    $r_0=(Z\alpha m)^{-1}$ is the Bohr radius.  The
    $\Upsilon$-potential comes into the equation of motion with the
    factor $w$, see the last term in Eq.(\ref{wa}).  Eq.(\ref{simp})
    gives an estimate $w\sim (mr_0)^{-1}|\boldsymbol{W}|\sim Z\alpha
    |\boldsymbol{W}|$.  Overall, an estimate for the correction
    produced by the $\Upsilon$-term in Eq.(\ref{wa}) is $\sim
    (Z\alpha)^3$, which means that the $\Upsilon$-term is too small to
    contribute to Eq.(\ref{dH}).  Thus, the contact and quadrupole
    interactions in Eq.(\ref{dH}) have no direst connection with the
    $\Upsilon$-term.  This fact makes a difference in coefficients in
    front of the contact term in Eq.(\ref{dH}) and the $\Upsilon$-term
    in Eq.(\ref{Ups}) acceptable. In particular, the fact that they
    have opposite signs produces no contradiction.

    \section{Coulomb problem}
    \label{Coulomb problem}

    Consider the pure Coulomb field, presuming that it is created by a
    point-like heavy object with charge $Z>0$. Then for $r>0$ one has
    \begin{eqnarray}
      \label{UF}
      U=-\frac{Z\alpha}{r}~,\quad \Upsilon=0~.
    \end{eqnarray}
    The second identity here follows from Eq.(\ref{Ups}).
    
    \subsection{Perturbation theory}
    
    Let us treat the Coulomb problem using the non-relativistic
    perturbation theory. Take the non-relativistic Eq.(\ref{nonrel})
    as a starting point, and consider the Hamiltonian Eq.(\ref{dH}) as
    a perturbation.  Conventional calculations, see Appendix
    \ref{appendix}, lead to the following result for the shift of the
    energy level characterized by the main quantum number $n$, orbital
    momentum $l$ and total angular momentum $j=l,l\pm 1$
    \begin{eqnarray}
      \label{nlj}
      \delta E_{nlj}= \frac{m(Z\alpha)^4}{n^3}\left(
        \frac{3}{8n}-\frac{1}{2j+1} \right)~.
    \end{eqnarray}
    This formula is similar to the one that describes the energy
    shifts for spinor particles; the only distinction comes from
    values of $j$ in Eq.(\ref{nlj}), which are integers for vector
    particles and half-integers for spinors.  The order of several
    lowest levels shows the following pattern
    \begin{eqnarray}
      \label{pattern}
      \begin{array}{llllll}
        n=1 &~~1 s_1\,; &        &        &       & 
        \\
        n=2 &~~2p_0,  &~~\{2s_1, & ~~2p_1\},  &~~2p_2\,;  &
        \\
        n=3 &~~3p_0,  &~~\{3s_1, & ~~3p_1,    &~~3d_1\},& 
        \\
            &        &   &~~\{3p_2, & ~~3d_2\},  &~~3d_3~.

      \end{array}
    \end{eqnarray}
    Here the atomic-like notation $nl_j$ is adopted, the brackets
    combine together the degenerate energy levels.

    \subsection{Central field}
    \label{central}    
    
    Consider the static central electric field (the Coulomb problem
    gives an important example). The conservation of the total angular
    momentum $j$ in this field allows one to separate the angular
    variables. We will use for this purpose the electric, longitudinal
    and magnetic spherical vectors, $\boldsymbol{
      Y}^{(e)}_{jm},\boldsymbol{ Y}^{(l)}_{jm}, \boldsymbol{
      Y}^{(m)}_{jm}$ defined conventionally, see \cite{LL4} and
    Appendix \ref{spherical}.  Generically, one can present the vector
    $\bf W$ as a linear combination of three spherical vectors with
    the given value of $j$. It is convenient to refer to the three
    terms in this combination as the electric, longitudinal and
    magnetic modes (or polarizations) of a vector boson. The parity
    conservation simplifies the problem further on. The state with the
    magnetic polarization, which parity is different from the parity
    of other two modes, is not coupled with these modes. Therefore the
    magnetically polarized mode can be written in a simple form
    \begin{eqnarray}
      \label{linm}
      \boldsymbol{ W}= f \,\boldsymbol{ Y}^{(m)}_{jm}~,
    \end{eqnarray}
    where $f=f(r)$ is the radial function. The two modes related to
    electric and longitudinal polarizations have same parity, which
    makes coupling between these modes possible. One needs therefore
    to consider them on the same footing assuming that
    \begin{eqnarray}
      \label{linel}
      \boldsymbol{ W}= u \,\boldsymbol{ Y}^{(e)}_{jm}+v \,\boldsymbol{ Y}^{(l)}_{jm}~,
    \end{eqnarray}
    where $u=u(r),\,v=v(r)$. We will refer to them as
    electro-longitudinal modes, or polarizations.
  
    \subsection{Magnetic polarization, $j\ge 1$}
    \label{magnetic}
    For the magnetic mode the angular momentum is restricted $j\ge 1$
    (the magnetic spherical vector is not defined for $j=0$, see
    Eq.(\ref{Y})). Substituting Eq.(\ref{linm}) into Eq.(\ref{wa}) one
    finds the following equation for the radial function $f$
    \begin{eqnarray}
      \label{wCoulomb}
      \big( \Delta_j+(\varepsilon+
          Z\alpha/r)^2-m^2 \big)\,f=0~.
    \end{eqnarray}
    Here $\Delta_j$ is 
    \begin{eqnarray}
      \label{Dj}
      \Delta_j= \frac{1}{r^2} \frac{d}{dr} \left(r^2\frac{d}{dr}\right)
        -\frac{j(j+1)}{r^2} ~.
      \end{eqnarray}
      The form of Eq.(\ref{wCoulomb}) coincides with the Klein-Gordon
      equation.  Therefore the spectrum of the magnetic mode
      replicates the spectrum of scalar particles, which is given by
      the Sommerfeld formula
    \begin{eqnarray}
      \label{Mspe}
      \varepsilon=m\left(1+\frac{(Z\alpha)^2}{\left(\gamma+n-j-1/2\right)^2}
      \right)^{-1/2}~.
    \end{eqnarray}
    Here
    \begin{eqnarray}
      \label{gam}
      \gamma=\left( \left( j+1/2\right)^2-(Z\alpha)^2 \right)^{1/2} ~.
    \end{eqnarray}
    In Eq.(\ref{Mspe}) $n=1,2 \dots$ plays a role of the main quantum
    number. In the non-relativistic limit the magnetic mode corresponds
    to the states $2p_1,3d_2,4f_3,...$.

    \subsection{Electro-longitudinal polarizations, $j\ge 1$}
    \label{j>0}
    Consider electro-longitudinal polarizations, when the vector $\bf
    W$ is given by Eq.(\ref{linel}). Substituting it into
    Eqs.(\ref{wa}),(\ref{simp}) and using the properties of the
    spherical vectors from Appendix \ref{spherical} one finds a system
    of coupled equations for radial functions $u,v$
    \begin{eqnarray}
      \label{sys1}
      \big(  \Delta_j+(\varepsilon
    +Z\alpha/r )^2-m^2 \big)u=
            -2\sqrt{j(j+1)}\,\,\frac{v}{r^2}&,&\quad\quad
      \\       \label{sys2}
      \big( \Delta_j+(\varepsilon+Z\alpha/r)^2-m^2 \big)v=
      -2\sqrt{j(j+1)}\,\,\frac{u}{r^2}&&\quad\quad
      \\ \nonumber
        +\frac{2v}{r^2}-
      \frac{2Z\alpha\,w}{r^2}~.&&
    \end{eqnarray}
    Here $w=w(r)$ denotes the radial part of ${\mathsf w}$.  Using
    Eq.(\ref{simp}) one finds for it
    \begin{eqnarray}
      \label{divW}
      &&{\mathsf w}=w\,Y_{jm}~,
      \\      \label{divWrad}
      &&w =
      \frac{1}{\varepsilon+Z\alpha/r}
      \left(-\sqrt{j(j+1)}\,\,\frac{u}{r}+\frac{dv}{dr}+\frac{2v}{r}\right).~~\quad
    \end{eqnarray}
    Eqs.(\ref{sys1}),(\ref{sys2}) are sufficient to define the
    functions $u,v$, but it is convenient to compliment them by the
    radial form of Eq.(\ref{0}), which reads
    \begin{eqnarray}
      \label{W0}
      \big( \Delta_j+(\varepsilon+Z\alpha/r)^2-m^2 \big) w =
      \frac{2 Z\alpha v}{r^2}~.
    \end{eqnarray}
    Let us verify first that Eqs.(\ref{sys1}),(\ref{sys2}) describe
    two different modes. Consider with this purpose distances so small
    that $m \ll Z\alpha/r$, where the potential energy dominates over
    mass. In this region Eqs.(\ref{sys1}),(\ref{sys2}) reduce to
    \begin{eqnarray}
      \label{sim1}
       &&\left( \frac{d^2}{dr^2}+\frac{2}{r}\frac{d}{dr}+
          \frac{(Z\alpha)^2-j(j+1)}{r^2}\right) u     \\
      \nonumber
       && 
       \quad\quad\quad\quad\quad\quad\quad\quad\quad~
       + 2 \left( j(j+1)\right)^{1/2}\frac{v}{r^2}=0,\\
      \label{sim2}
      &&\left(\frac{d^2}{dr^2}+\frac{4}{r}\frac{d}{dr}+
        \frac{(Z\alpha)^2-j(j+1)+2}{r^2}\right) v=0.\quad\quad
    \end{eqnarray}
    One derives from Eqs.(\ref{sim1}),(\ref{sim2}) that there exists a
    mode, in which at small distances $v$ is small, $|v|\ll |u|$,
    which means that in this region the polarization is predominantly
    electric.  From Eq.(\ref{sim1}) one finds that this mode satisfies
    the following asymptotic conditions at $r\rightarrow 0$
    \begin{eqnarray}
      \label{uas}
      u \rightarrow a \, r^{\gamma-1/2}~,\quad |v| \ll |u|~.
    \end{eqnarray}
    We will call it the ``$\gamma-1/2$'' mode below.

    In order to find the second mode let us assume the following
    asymptotic behavior for $r\rightarrow 0$
    \begin{eqnarray}
      \label{anu}
      &&u\rightarrow b\,r^\nu~,\\
      \label{bnu}
      &&v\rightarrow c\,r^\nu~,
    \end{eqnarray}
    Substituting Eqs.(\ref{anu}),(\ref{bnu}) in
    Eqs.(\ref{sim1}),(\ref{sim2}) one finds a system of two
    homogeneous linear equations, in which $\nu$ plays a role of
    the eigenvalue. Solving this system one finds $\nu$ and the ratio
    $c/b$, deriving
    \begin{eqnarray}
      \label{nu}
      && u\rightarrow  b\,\,r^{\gamma-3/2}~,\\
      \label{banu}
      && v\rightarrow b\,\frac{\gamma-1/2 }{ \sqrt{ j(j+1)} } \,\,\,
         r^{\gamma-3/2}.~
    \end{eqnarray}
    This mode will be referred to as the ``$\gamma-3/2$'' mode \cite{l=jpm1}.
    
    Let us find now the discrete energy spectrum.  Introduce a
    function $g=g(r)$
    \begin{eqnarray}
      \label{chi}
      g&=&Z\alpha u+\sqrt{j(j+1)}\,w
      \\       \nonumber 
      &=&Z\alpha u
      +\frac{ \sqrt{j(j+1)} }{ \varepsilon+Z\alpha/r }
        \left(-\sqrt{j(j+1)} \,\frac{u}{r}+\frac{dv}{dr}
        +\frac{2v}{r}\right)~. 
    \end{eqnarray}
    Here Eq.(\ref{divWrad}) was used in the second identity. Taking
    the corresponding linear combination of
    Eqs.(\ref{sys1}),(\ref{W0}) one finds that $g$ satisfies the
    Klein-Gordon equation
    \begin{eqnarray}
      \label{Dchi}
      \big( \,\Delta_j+(\varepsilon+Z\alpha/r)^2-m^2\,
      \big)
      g=0~.
    \end{eqnarray}
    This result leaves only two options; either $g$ equals zero
    identically, or, alternatively, the spectrum of
    electro-longitudinal modes can be found from Eq.(\ref{Dchi}).  The
    first alternative takes place for $j=0$, when only the
    longitudinal mode is present. The function $u$ in this case should
    be taken as zero, which makes zero also the function $g$ in
    Eq.(\ref{chi}).  Thus, Eq.(\ref{Dchi}) provides no help for $j=0$
    states.
    
    For $j\ge 1$ the function $g$ is nonzero, for both
    ``$\gamma-1/2$'' and ``$\gamma-3/2$'' modes, see Appendix
    \ref{non0}. Eq.(\ref{Dchi}) defines the spectrum, which therefore
    satisfies the Sommerfeld formula Eq.(\ref{Dchi}).  In the
    non-relativistic limit the mixed electric-longitudinal modes
    correspond to the following states with $j\ge 1$: $1s_1, 2p_2,
    3d_1, 3d_3, 4f_2 \ldots$.

    \subsection{Longitudinal polarization, $j=0$}
    \label{j=0}
    
    Consider zero angular momentum $j=0$, which corresponds to purely
    longitudinal polarization, see Eq.(\ref{Y}).  The state with $j=0$
    is described by one radial function $v=v(r)$,
    \begin{eqnarray}
      \label{pres}
      \boldsymbol{ W}= v \,\boldsymbol{n}~,\quad\quad
      \boldsymbol{n}=\boldsymbol{r}/r~.
    \end{eqnarray}
    The radial function $v$ satisfies Eqs.(\ref{sys1}),(\ref{sys2}) in
    which the function $u$ is to be put to zero (electric polarization
    for $j=0$ is impossible). These equations therefore yield
    \begin{eqnarray}
      \label{v}
      \frac{d^2v}{dr^2}+\frac{2}{r}\,\frac{dv}{dr}&+&\left(
        \left(\varepsilon +Z\alpha/r\right)^2-m^2
      \right)v\\
      \nonumber 
      & =& \frac{2v}{r^2}
             -\frac{2Z\alpha}{r^2}\,
       \frac{1}{\varepsilon+Z\alpha/r} \left(\frac{dv}{dr}
         +\frac{2v}{r}\right)~.
    \end{eqnarray}
    In order to make the physical meaning of this equation more
    transparent let us eliminate the first derivative by means of a
    substitution $v\rightarrow \varphi$
    \begin{eqnarray}
      \label{phin}
      v =\frac{Z \alpha} {\varepsilon^2} \,\left( 
        \varepsilon+\frac{Z\alpha}{r}\right)\,
      \frac{\varphi}{r}=
      \frac{1+x}{x^2}\, \varphi~.
    \end{eqnarray}
    where it is convenient also to scale the radial variable
    $r\rightarrow x$
    \begin{eqnarray}
      \label{x}
      r=\frac{Z\alpha}{\varepsilon}\,x~,
    \end{eqnarray}
    assuming $\varphi=\varphi(x)$.  In this notation Eq.(\ref{v}) can
    be rewritten as a conventional Schr\"odinger-type eigenvalue
    problem
    \begin{eqnarray}
      \label{schr}
&&      H\,\varphi=-\varkappa^{\,2}\varphi~, \\
      \label{H}
&&      H=-\frac{d^2}{dx^2} 
      -\frac{2(Z\alpha)^2}{x}
        -\frac{(Z\alpha)^2}{x^2}+\frac{2}{(x+1)^2}~,\quad\quad
    \end{eqnarray}
    where $-\varkappa^2$, which plays a role of an eigenvalue, is
    related to the energy of the discrete level
    \begin{eqnarray}
      \label{kappa}
      \varkappa^{\,2}=(Z\alpha)^2\frac{m^2-\varepsilon^2}{\varepsilon^2}>0~.
    \end{eqnarray}
    The operator $H$ in Eq.(\ref{H}) possesses three singular points,
    $x=0,~x=\infty$ and $x=-1$. The last one lies in the non-physical
    region, but it presents an obstacle for an analytical study
    anyway.  One can overcome this difficulty using a substitution
    $\varphi \rightarrow \tilde \varphi$
    \begin{eqnarray}
      \label{for}
            \varphi=\left( \frac{d}{dx}
        +\left(\gamma+1/2\right)\frac{x+1}{x}-\frac{1}{x+1}
      \right)\tilde \varphi~.
    \end{eqnarray}
    It can be shown that $\tilde \varphi$ satisfies an eigenvalue
    problem
    \begin{eqnarray}
      \label{H2}
      &&\tilde H\tilde \varphi=-\varkappa^2\,\tilde \varphi~,
      \\ \label{psi1}
      &&\tilde H=
      -\frac{d^2}{dx^2}-2\frac{(Z\alpha)^2}{x}
   +\Big(\gamma+\frac{1}{2}\Big)\Big(\gamma+\frac{3}{2}\Big)\frac{1}{x^2}~.\quad
   \quad\quad
    \end{eqnarray}
    The main result of the transformation Eq.(\ref{for}) is that the
    operator $\tilde H$ has only two singular points, $x=0$ and
    $x=\infty$.  An interesting method, which allows one to ``invent''
    the substitution Eq.(\ref{for}) and derive then Eq.(\ref{H2}) is
    presented in Appendix \ref{matrix}. It takes its origins in an
    elegant treatment of quantum mechanics developed by the G\"otingen
    School and known as matrix mechanics.
    
    A regular at $r=0$ solution of the eigenvalue problem (\ref{H2})
    reads
    \begin{eqnarray}
      \label{psi2in}
      \tilde \varphi=e^{-\varkappa x}x^{L+1}
      F\left(L+1-\frac{(Z\alpha)^2}{\varkappa},2L+2,2\varkappa 
        x\right).\quad
    \end{eqnarray}
    Here $F(\alpha,\beta, z)$ is the confluent hypergeometric function
    and $L$ is defined by
    \begin{eqnarray}
      \label{L}
      L=\gamma+1/2~. 
    \end{eqnarray}
    To make the solution given by Eq.(\ref{psi2in}) regular at
    infinity one should assume that
    \begin{eqnarray}
      \label{Ry}
      \varkappa=
      \frac{(Z\alpha)^2}{L+n-1}=
      \frac{(Z\alpha)^2}{\gamma+n-1/2}
      ~,\quad n=2,3\ldots~,~~
    \end{eqnarray}
    The corresponding eigenfunctions are given by Eq.(\ref{psi2in}),
    in which the hypergeometric function is reduced to a polynomial
    \begin{eqnarray}
      \label{psi2}
    \tilde \varphi=e^{-\varkappa x}x^{\gamma+3/2} F(2-n,2\gamma+3,2\varkappa x
    )~. 
    \end{eqnarray}
    Eqs.(\ref{psi2}),(\ref{for}) give then the function $\varphi$,
    while (\ref{phin}),(\ref{pres}) transform it into $v$ and
    $\boldsymbol W$. The function $\varphi$ exhibits the following
    behavior at the boundaries
    \begin{eqnarray}
      \label{sens1}
      &\varphi \propto \exp(-\varkappa \,x)~,\quad &x\rightarrow \infty~,
      \\
      \label{sens2}
      &\varphi \propto x^{\gamma+1/2}~, \quad &x\rightarrow 0~.
    \end{eqnarray}
    Eq.(\ref{Ry}) gives the spectrum
    \begin{eqnarray}
      \label{sati}
      \varepsilon=m\left(1+\frac{(Z\alpha)^2}{\left(\gamma+n-1/2\right)^2}
      \right)^{-1/2}\!\!,\quad n=2,3\ldots\quad
    \end{eqnarray}
    which complies with the Sommerfeld formula Eq.(\ref{Mspe}). In the
    non-relativistic limit the longitudinal mode corresponds to the
    following states with $j=0$: $2p_0, 3p_0, 4p_0\ldots$
   
    \subsection{Summary for Coulomb problem}
    
    \label{Spectrum}
    
    Our discussion of the Coulomb problem for vector particles
    confirms that for all polarizations and all angular momenta $j$
    the discrete energy spectrum is described by the Sommerfeld
    formula Eq.(\ref{Mspe}), as was first found by Corben and
    Schwinger \cite{corben-schwinger_1940}.
    
    For $j\ge 1$ there exist three modes.  One of them is purely
    magnetic, it has $l=j$, while two others are constructed from the
    electric and longitudinal polarizations, each one of these two
    modes has an admixture of $l=j+1$ and $l=j-1$ states.  These two
    modes coexist for $j\ge 1$, while for $j=0$ only one of them,
    which in this case has a purely longitudinal polarization and
    $l=1$ is present. 
    
    From Eq.(\ref{Mspe}) one derives that the spectrum of the Coulomb
    problem is degenerate; it is triply degenerate provided $n \ge
    j+2,~j\ge 1$, doubly degenerate for levels with $n=j+1,~j\ge 1$,
    while the states which have either $n=j$ or $j=0$ remain
    non-degenerate.  This conclusion agrees with the non-relativistic
    expansion, see Eq.(\ref{pattern}).
    
    Interestingly, one and the same Sommerfeld formula Eq.(\ref{Mspe})
    describes the discrete energy spectrum in the Coulomb problem for
    scalar, Dirac and vector particles. The only distinction is
    related to the angular momentum $j$, which takes the integer
    values $j=0,1\ldots $ for bosons and half-integer $j=1/2,3/2\ldots
    $ for fermions.

    \section{catastrophe with charge}
    \label{catastroph}
    
    Consider the charge density of a vector boson for a state with
    $j=0$.  Eqs.(\ref{ro}),(\ref{pres}) give
    \begin{eqnarray}
      \label{rode}
           \rho^W= 2e\left[ \left(\varepsilon+\frac{Z\alpha}{r}\right)(v^2+w^2)
     +2v \frac{dw}{dr}-\Upsilon\, w^2\right],\quad
    \end{eqnarray}
    where $w$ defined by Eqs.(\ref{simp}),(\ref{divW}),(\ref{divWrad}) reads
    \begin{eqnarray}
      \label{wj0}
      w=\frac{1}{\varepsilon+Z\alpha/r} \left(\frac{dv}{dr}+
        \frac{2v}{r}\right)~.
    \end{eqnarray}
    In the region of small distances $r\ll Z\alpha/m$ Eq.(\ref{sens2})
    shows that $\varphi \propto r^{\gamma+1/2}$. Consequently, from
    Eqs. (\ref{phin}),(\ref{wj0}) we find the following estimates for
    $v$ and $w$
    \begin{eqnarray}
      \label{estv}
      v &\sim& r^{\gamma-3/2}~,
      \\ \label{estw}
      w &\sim& \frac{\gamma+1/2}{Z\alpha} r^{\gamma-3/2}~,
    \end{eqnarray}
    From here one derives an estimate for the charge density
    (\ref{rode}) of the vector boson
    \begin{eqnarray}
      \label{cdvb}
      \rho^W\sim -2e\,
      \frac{(1-\gamma)(1+2\gamma)}{Z\alpha}\,r^{2\gamma-4}~,
      \quad r>0~.
    \end{eqnarray}
    It diverges at the origin so badly that the total charge $Q^W=\int
    \rho^W\,d^3r$ localized in any small sphere surrounding the origin
    is infinite.
    
    The trouble does not stop here.  Remember the density $\rho=
    Z|e|\,\delta(\boldsymbol{r})$ of the Coulomb charge, which is
    located at the origin.  This density results in the
    $\Upsilon$-term defined by Eq.(\ref{Ups})
    \begin{eqnarray}
      \label{delta}
      \Upsilon= \frac{e\rho}{\,m^2} = 
      -\frac{4\pi Z\alpha}{m^2}\,\delta(\boldsymbol{r})~.
    \end{eqnarray}
    We did not consider it previously because the functions we dealt
    with were regular at the origin, allowing one to hope that their
    regular behavior makes the $\Upsilon$-term irrelevant. Since the
    charge density does not follow this pattern, we need to take the
    term given by Eq.(\ref{delta}) into account. The contribution of
    the $\delta$-function in Eq.(\ref{delta}) to the boson charge
    density is given by the last term in Eq.(\ref{rode}), which reads
    \begin{eqnarray}
      \label{ro4}
      (\rho^W)_{\,\Upsilon-\mathrm{term}}
      =e \frac{8\pi
      Z\alpha}{m^2}\,w^2(0)\,\delta(\boldsymbol{r}) ~.
    \end{eqnarray}
    Eq.(\ref{estw}) shows that $w(0)=\infty$, which makes the density
    Eq.(\ref{ro4}) infinite as well.
    
    We see that there are two closely located, though different
    regions, which contribute to an infinite charge of the $W$-boson
    in the $j=0$ state. One region is $r>0$, where the density of
    charge Eq.(\ref{cdvb}) behaves singularly as $r\rightarrow 0$.
    Another region is located strictly at the origin $r=0$, where an
    infinite coefficient $w^2(0)=\infty$ in front of the
    $\delta$-function in Eq.(\ref{ro4}) makes the charge infinite as
    well.
        
    The origin of Eq.(\ref{ro4}) can be traced down to the last term
    in Eq.(\ref{ro}). It contributes therefore to the charge density
    for all states.  There is one more state, in which the coefficient
    in Eq.(\ref{ro4}) turns infinite, signaling a catastrophic
    behavior of the charge.  This is the ``$\gamma-3/2$'' state with
    $j=1$, see Eqs.(\ref{nu}),(\ref{banu}).  To justify this
    statement, note that Eqs.(\ref{divWrad}),(\ref{nu}) and (\ref{banu})
    imply that $w\propto r^{\gamma-3/2}$.  For $j=1$ the inequality
    $\gamma< 3/2$ holds.  Therefore for the state ``$\gamma-3/2$'',
    $j=1$ one finds $w(0)=\infty$, which makes the charge of the
    $W$-boson located strictly at the origin infinite.  (There is no
    problem in that case with the charge in the region $r>0$.)
    
    The catastrophic behavior of the charge of the $W$-boson in $j=0$
    and $j=1$, ``$\gamma-3/2$'' states was discovered in
    \cite{corben-schwinger_1940}, forcing the authors of this work to
    conclude that the pure Coulomb problem for $W$-bosons is poorly
    defined.

    \section{Vacuum polarization}
    \label{vacuum_polarization}

    Consider the conventional QED vacuum polarization. The potential
    energy of the $W$-boson propagating in the Coulomb field acquires
    an additional term, let us call it $S(r)$, which describes the
    polarization
      \begin{eqnarray}
        \label{pot}
        U(r)=-\Big(\,1+S(r)\,\Big) \frac{Z\alpha}{r}~.
      \end{eqnarray}
      It suffices to consider the polarization effect in the
      lowest-order approximation, when it is is described by the known
      Uehling potential. Its small-distance asymptotic behavior is
      given by a simple logarithmic function, see e. g. \cite{LL4},
      \begin{eqnarray}
        \label{write}
        S(r)\simeq -\alpha \beta \,\ln\left(m_{Z}r\right),
        \quad
        r\rightarrow 0~.
      \end{eqnarray}
      This function is related to the logarithm responsible for the
      scaling of the QED coupling constant
      \begin{eqnarray}
        \label{log}
        \alpha^{-1}(\mu)=\alpha^{-1}(\mu_0)-\beta\,\ln(\mu/\mu_0)~. 
      \end{eqnarray}
      The relation between Eqs.(\ref{write}) and (\ref{log}) is
      well-known, see e.g. book \cite{LL4}, which presents it for one
      generation of leptons. The factor $\beta$, which governs the
      scaling of the coupling constant and the potential in
      Eq.(\ref{write}) equals the lowest coefficient of the Gell-Mann
      - Low $\beta$-function.  It is normalized here in such a way
      that for one generation of leptons $\beta=\beta_{e}=2/3\pi$.

      It is important for us that $\alpha(\mu)$ rises with the mass
      parameter $\mu$, i.e. $\beta$ is positive, $\beta>0$;
      theoretical and experimental data agree on this fact, for a
      brief review see e. g. Ref. \cite{eidelman-et-al_2004}, the
      experimental data are provided by Refs. \cite{TOPAZ,VENUS,OPAL}.
      An estimation of $\beta$ can be found from two reliable
      reference points $ \alpha^{-1}(m_\tau) = 133.498 \pm 0.017$ and
      $\alpha^{-1}(m_Z) = 127.918 \pm 0.018$ provided in
      Ref.\cite{eidelman-et-al_2004}.  Using them and taking the
      masses $m_\tau= 1776.99 + 0.29 - 0.26$ Mev and $m_Z= 91.1876 \pm
      0.0021$ Gev recommended in \cite{eidelman-et-al_2004} one
      derives from Eq.(\ref{log}) that
      \begin{eqnarray}
        \label{bet}
        \beta \simeq 1.42(1)~. 
      \end{eqnarray}
      More simple estimation of $\beta$ can be done if one takes into
      account a contribution of all known charged fermions ``naively''
      (neglecting complications, related to the QCD vacuum as well as
      possible contribution of scalars).  This estimate yields
      \begin{eqnarray}
        \label{betr}
        \beta_\mathrm{est} \approx \frac{2}{3\pi}\sum_i \frac{q^2_i}{e^2}
        =\frac{2}{3\pi}\left(1+\frac{5}{3}\right)3\simeq 1.70~. 
      \end{eqnarray}
      Here summation runs over all charged fermions, $q_i$ is the
      charge of the fermion, the terms 1 and 3/5 in the bracket are
      due to the lepton and quark contribution for one generation, the
      factor 3 after the bracket accounts for three generations.  A
      discrepancy between ``simple-minded'' Eq.(\ref{betr}) and more
      solid-based Eq.(\ref{bet}) is below 20\%.  The normalization of
      the logarithmic function on the mass of the $Z$-boson $m_Z$
      adopted in Eq.(\ref{write}) presumes that the fine-structure
      constant $\alpha $ is taken at precisely this scale,
      $\alpha\equiv\alpha(m_Z)\simeq 1/128$.

      We are interested in high-momenta behavior in
      Eqs.(\ref{write}),(\ref{log}), where $\mu\sim 1/r \gg m$.  An
      accuracy of Eqs.(\ref{bet}),(\ref{betr}), as well as any other
      feasible estimation, is limited in this region by a contribution
      of unknown heavy charged fermions and scalars. However, this
      uncertainty does not affect our final conclusions.  For our
      purposes it suffices to stick to a widely accepted hypothesis
      that $\beta$ is a positive constant (or a slow-varying function
      up to the Grand Unification limit).

      Substituting Eqs.(\ref{pot}),(\ref{write}) into Eq.(\ref{Ups})
      one derives
      \begin{eqnarray}
        \label{estUps}
        \Upsilon(r)&\simeq& 
        \frac{Z\alpha^2 \beta}{m^2r^3},\quad
        r\rightarrow 0~,
      \end{eqnarray}
      where the lowest term of the $\alpha$-expansion is retained.  It
      is vital that for small distances, when $r\ll \sqrt{\alpha}/m$,
      $\Upsilon(r)$ is positive and large,
      \begin{eqnarray}
        \label{large}
        \Upsilon(r) \gg |U(r)|\gg m~.
      \end{eqnarray}
      Note that the direct contribution of the vacuum polarization
      given by the term $S(r)$ in Eq.(\ref{pot}) is not pronounced. In
      contrast, the $\Upsilon$-term Eq.(\ref{estUps}) becomes dominant
      at small distances, making the effects related to the QED vacuum
      polarization very important.  Since this term plays a crucial
      role below, let us verify its sign again.  Consider a positive
      Coulomb center, $Z>0$.  Then the vacuum polarization produces
      negative charge density, $\rho <0 $. Since the charge of the
      $W^-$ meson is negative, $e<0$, we find from Eq.(\ref{Ups}) that
      $\Upsilon =e \rho/m^2>0$.  We see that indeed, the
      $\Upsilon$-term is positive, in accord with Eq.(\ref{estUps}).

    \subsection{Longitudinal polarization, j=0}
    \label{longitudinal_polarization_j=0}
        
    Eq.(\ref{pres}) shows that a longitudinal state with $j=0$ is
    described by the single radial function $v=v(r)$.  Eq.(\ref{simp})
    allows one to express the function $w$ via $v$
      \begin{eqnarray}
        \label{wv}
      w= \left(\varepsilon-U-\Upsilon\right)^{-1}\,\big(v'+2v/r\big)~.
    \end{eqnarray}
    We need now to write the classical equation of motion for $v$, in
    which the term $\Upsilon$ is taken into account.  The simplest way
    is to substitute $\boldsymbol{W}$ and $w$ from
    Eqs.(\ref{pres}),(\ref{wv}) into Eq.(\ref{tran}), which yields
      \begin{eqnarray}
        \label{radv}
    &&   \big((\varepsilon-U)^2\!-m^2\big)v=
    \\ \nonumber
   && -(\varepsilon-U)\frac{d}{dr} \left(\,
    \frac{v'+2v/r }{\varepsilon-U-\Upsilon}
       \,\right)
    -U'\,\frac{v'+2v/r}{\varepsilon-U-\Upsilon}.
    \end{eqnarray}
    It is taken into account here that Eq.(\ref{pres}) ensures that
    $\boldsymbol{\nabla \!\times \!W}= 0$. For a purely Coulomb case,
    when $\Upsilon=0$ for $r>0$, Eq.(\ref{radv}) reduces to
    Eq.(\ref{v}).  Eq.(\ref{radv}) can be rewritten in a more compact
    form
      \begin{eqnarray}
        \label{sec}
        v''+G \, v'+H \,v=0~,
      \end{eqnarray}
      where the coefficients $G=G(r)$ and $H=H(r)$ are
      \begin{eqnarray}
        \label{g}
        G &=& \frac{2}{r}+\frac{U'}{\varepsilon-U}+\frac{U'+\Upsilon'}
        {\varepsilon-U-\Upsilon}~,
        \\ \label{h}
        H &=&-\frac{2}{r^2}+\frac{2}{r}
        \left( \frac{U'}{\varepsilon-U}+\frac{U'+\Upsilon'}{\varepsilon-U-\Upsilon}
        \right)
        \\ \nonumber
        &&\quad \quad \quad +\frac{\varepsilon-U-\Upsilon}{\varepsilon-U}
        \big( \, (\varepsilon-U)^2-m^2 \,\big)~.
      \end{eqnarray}
      For a qualitative analyses it is convenient to eliminate the
      term with the first derivative by scaling the radial function
      $v\rightarrow\varphi=\varphi(r)$
      \begin{eqnarray}
        \label{phi}
        v=\frac{1}{r}\Big[ \, \big(\varepsilon-U\,\big)
        \big(\varepsilon-U-\Upsilon\,\big)
        \,\Big]^{1/2}\,\varphi~.
      \end{eqnarray}
      (This definition reduces to Eq.(\ref{phin}) when $\Upsilon=0$).
      The classical equation of motion for $W$-bosons takes a simple
      form
      \begin{eqnarray}
        \label{phi''}
       && -\varphi''+{\mathcal U}\,\varphi=0~,
        \\ \label{ugh}
        &&{\mathcal U}=-\,H+G^{\,2}/4+G{\,'}/2~,
      \end{eqnarray}
      where $G,H$ are defined in Eqs.(\ref{g}),(\ref{h}).
      Eq.(\ref{phi''}) can be looked at as a Schr\"odinger-type
      equation, in which ${\mathcal U}={\mathcal U}(r)$ plays the role of an
      effective potential energy.

      According to Eqs.(\ref{estUps})(\ref{large}) the $\Upsilon$-term
      is large and positive at small distances. This fact makes the
      effective potential ${\mathcal U}(r)$ in Eq.(\ref{ugh}) also large and
      positive when $r\rightarrow 0$
         \begin{eqnarray}
        \label{larpos}
        {\mathcal U}(r)\simeq -\,H(r)\simeq -\,U(r)\,\Upsilon(r)\simeq
          \frac{Z^2\alpha^3\beta}{m^2r^4}~.
      \end{eqnarray}
      Compare this result with the effective potential $ [ \,{\mathcal
        U}(r)\,]_\mathrm{C}$ for the pure Coulomb field.  The latter one is a
      part of the Hamiltonian in Eq.(\ref{H}).  For $r\rightarrow
      0$ one finds from Eq.(\ref{schr}) that
      \begin{eqnarray}
        \label{UC}
       [ \,{\mathcal U}(r)\,]_\mathrm{C}\simeq -(Z\alpha)^2/r^2.
      \end{eqnarray}
      It is taken into account here that the variables $x$ and $r$ in
      Eq.(\ref{H}) are proportional, see Eq.(\ref{x}).
      Eq.(\ref{larpos}) shows that the vacuum polarization produces a
      strong repulsion in the effective potential ${\mathcal U}(r)$, in
      contrast with a mild attraction, which exhibits $[\,{\mathcal
        U}(r)\,]_\mathrm{C}$ in Eq.(\ref{UC}) for the pure Coulomb
      case.
      
      When the estimate Eq.(\ref{larpos}) is applicable,
      Eq.(\ref{phi''}) allows an analytical solution
      \begin{eqnarray}
        \label{re}
        \varphi(r)\propto 
      r \,\exp\left(\!-\frac{Z
        \alpha\,(\alpha \beta)^{1/2}}{mr} \,\right).~~
      \end{eqnarray}
      It shows that $\varphi(r)$ exponentially decreases at small
      distances.  According to Eqs.(\ref{wv}),(\ref{phi}) the
      functions $v(r),w(r)$, also decrease exponentially here;
      correspondingly, the charge density of the $W$-boson
      Eq.(\ref{rode}) decreases exponentially at the origin as well
      \begin{eqnarray}
        \label{vexp}
        v&\rightarrow& \frac{a}{m}\,\frac{1}{r^2}
          \exp\left(\!-\frac{Z
        \alpha\,( \alpha \beta)^{1/2}}{mr} \,\right)~,
    \\
        \label{wexp}
        w &\rightarrow &
        -\frac{a}{(\alpha \beta)^{1/2} }\,\frac{1}{r}
          \exp\left(\!-\frac{Z
        \alpha\,(\alpha \beta)^{1/2}}{mr} \,\right)~,
    \\
        \label{rhoexp}
        \rho^W &\rightarrow& 
        -\frac{4 a^2 Z\alpha \,e}{m^2 }\,\frac{1}{r^5}
          \exp\left(\!-\frac{2Z
        \alpha\,(\alpha \beta)^{1/2}}{mr} \,\right).\quad
      \end{eqnarray}
      Here a constant $a$ depends on the normalization of $v$, which
      is specified in Eq.(\ref{norma}) below.  Eq.(\ref{rhoexp}) shows
      that for $r>0$ in the vicinity of the origin the charge density
      is finite and small; which makes the charge located in this
      region finite as well.  Eq.(\ref{wexp}) shows that $w(0)=0$,
      which eradicates the contribution of the $\delta$-function in
      Eq.(\ref{ro4}). Thus, the charge located strictly at origin
      $r=0$ is zero.
      
      We verified that an account of the QED vacuum polarization
      erases the infinite charge of a vector boson for $j=0$ state.

      \subsection{Electro-longitudinal polarizations, $j\ge 1$}
      \label{Electro-longitudinal polarizations, j>0}
      
      Eq.(\ref{large}) shows that in the region of small distances
      $r\ll \alpha/m$ the $\Upsilon$-term, which is related to the
      vacuum polarization, is large. This fact makes the function $w$
      in Eq.(\ref{simp}) small, $|w|\ll |\boldsymbol{W}|$.  As a
      result the asymptotic form of the equation of motion
      (\ref{tran}) at small distances reads
    \begin{eqnarray}
      \label{tran1}
       \frac{(Z\alpha)^2\!\!}{r^2}\,\,\boldsymbol{W}=
    \boldsymbol{\nabla \times}(\boldsymbol{
      \nabla \times W})~.
      \end{eqnarray}
      Eq.(\ref{linel}) ensures that $ \boldsymbol{ \nabla \times W}$
      is not zero identically  provided $j\ge 1$, which makes
      Eq.(\ref{tran1}) meaningful. 
      
      Using Eq.(\ref{linel}) to represent the electro-longitudinal
      modes and identities Eqs.(\ref{rot}) for the spherical vectors
      one rewrites Eq.(\ref{tran1}) in terms of the radial functions
      $u,v$
      \begin{eqnarray}
        \label{rad u}
       (Z\alpha)^2u=-r^2 u''-r u' -u+ \sqrt{j(j+1)}\,v\,, \quad\quad\,    &&
       \\ \label{rad v}
 (Z\alpha)^2 v=-\sqrt{ j(j+1)}\,( \,r u' + u-
       \sqrt{j(j+1)}\,v \,).&&\quad       
      \end{eqnarray}
      Their solution is straightforward
      \begin{eqnarray}
        \label{sol u}
       u&=& b \,r^{\lambda}~,
       \\ \label{sol v}
       v&=&b \,\sqrt{
       j(j+1)}\,\frac{\lambda+1}{\gamma^2-1/4}\,r^{\lambda}~.
      \end{eqnarray}
      Here $b$ is a constant, and $\lambda$ can take one of the two
      possible values, 
      \begin{eqnarray}
        \label{lpm}
        \lambda=\lambda_\pm=\frac{1}{2}\,\,\frac{j(j+1)\pm k}
        {\gamma^2-1/4}~,
      \end{eqnarray}
      where $k$ satisfies
      \begin{eqnarray}
        \label{k}
        k^2= j^2(j+1)^2
          -4(Z\alpha)^2\left(\gamma^2-\frac{5}{4}\right)\left(\gamma^2
            -\frac{1}{4}\right),\quad
      \end{eqnarray}
      with $\gamma$ defined in Eq.(\ref{gam}).  The two available
      values of $\lambda_\pm$ should be attributed to the two
      electro-longitudinal modes.
            
      Comparing Eqs.(\ref{sol u}),(\ref{sol v}), which are valid when
      the vacuum polarization is taken into account, with
      Eqs.(\ref{uas}) and Eqs.(\ref{nu}),(\ref{banu}), which describe
      the purely Coulomb case, we see that the polarization changes
      drastically the behavior of the wave functions. One finds from
      Eqs.(\ref{lpm}),(\ref{k}) that $\lambda_\pm$ are positive for
      all $j$, $j\ge 1$, provided $Z$ is not very large, $Z\alpha\le
      1/2$. 
      
      From Eqs.  (\ref{sol u}),(\ref{sol v}) one deduces therefore
      that $u(r),v(r)\rightarrow 0$, when $r\rightarrow 0$.
      Eq.(\ref{simp}) guarantees then that $w(0)=0$. As a result the
      contribution of the $\delta$-function in Eq.(\ref{ro4}) to the
      charge density turns zero for all electro-longitudinal modes.
      This differs qualitatively from the pure Coulomb case, which
      gives an infinite charge located at the origin for $j=1$,
      ``$l$''$=0$ state.
      
      We conclude that the QED vacuum polarization suppresses the wave
      functions of a vector boson at the origin, eradicating thus the
      infinite charge of the boson, which plagues the problem for the
      pure Coulomb field.

      \section{Numerical example}
      \label{numericals}
      To be more informative on the behavior of vector bosons in the
      Coulomb field let us solve the corresponding equations of motion
      numerically. Consider the $j=0$ state, describing it with the
      help of Eqs.(\ref{pres}) and (\ref{radv}). We need to specify
      the factor $S(r)$, which describes the vacuum polarization in
      the potential in Eq.(\ref{pot}).  For small $r$, $r\ll
      Z\alpha/m$ this factor plays a major role, while for larger it
      is less important.  Let us construct a simple model, which gives
      a correct asymptotic behavior Eq.(\ref{write}) as $r\rightarrow
      0$, and is physically reasonable, though not perfect, at larger
      $r$. Take with this purpose the Uehling potential, see e.g.
      \cite{LL4}, assuming that only charged leptons and quarks
      contribute to it
      \begin{eqnarray}
        \label{Ueh}
        S(r)=\frac{2\alpha}{3\pi}\sum_i \,\frac{q_i^2}{e^2}\, F(m_ir)~.
      \end{eqnarray}
      Here
      \begin{eqnarray}
       \label{F(x)}
        F(x)= \int_1^\infty \!\!\exp(-2x\zeta) \left( 1+\frac{1}{2\zeta^2}\right)
          \frac{\sqrt{\zeta^2-1}}{\zeta^2}\,\,d\zeta~.
      \end{eqnarray}
      Summation in Eq.(\ref{Ueh}) runs over all quarks and charged
      leptons, their charges $q_i$ and masses $m_i$ are taken from
      Ref. \cite{eidelman-et-al_2004}. The model presented by
      Eq.(\ref{Ueh}) neglects complications related to the QCD vacuum,
      which may be substantial at large distances, but the role of the
      polarization is insignificant in this region anyway. For small
      distances $r\rightarrow 0$ the model Eq.(\ref{Ueh}) reproduces
      the correct asymptotic formula Eq.(\ref{write}), in which the
      coefficient $\beta$ is given by Eq.(\ref{betr}).  A comparison
      of Eqs.(\ref{bet}) and (\ref{betr}) shows that at small
      distances the accuracy of the model potential can be estimated
      as $\sim 20$\%, which is sufficient for us.
\begin{figure}
\centering
 \includegraphics[height=5.3 cm,
keepaspectratio = true, 
]{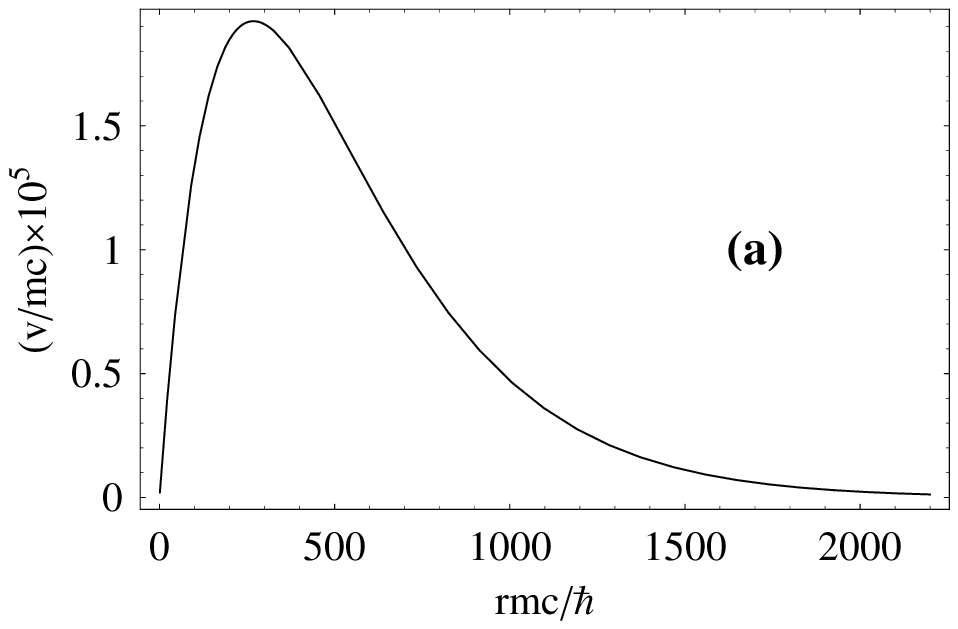}
\centering
 \includegraphics[height=5.3 cm,
keepaspectratio = true, 
]{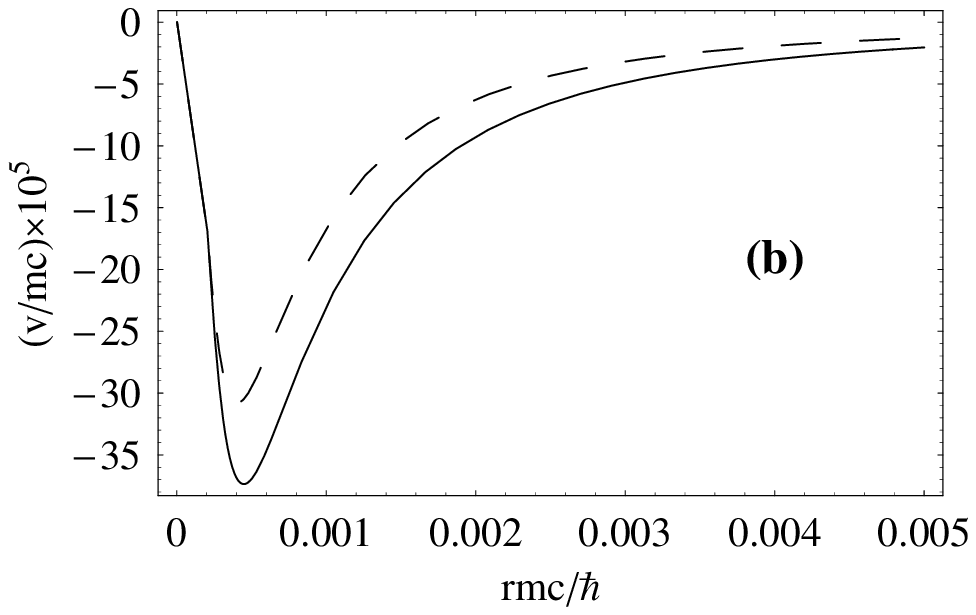}
\caption{ \label{one} The $2p_1$ radial function $v(r)$, which describes the
  $n=2$, $j=0$ discrete state of a $W$-boson in the Coulomb field of a
  charge $Z=1$. (a) Large distances $r\gg \hbar/mc$, $v(r)$ is close
  to conventional non-relativistic wave function $2p$; (b)
  ultra-relativistic region $r\ll \hbar/mc$.  Solid line - numerical
  solution, dashed line - analytical prediction of Eq.(\ref{vexp}). }
\end{figure}
     \noindent 
\begin{figure}
\centering
 \includegraphics[height=5.3 cm,
keepaspectratio = true, 
]{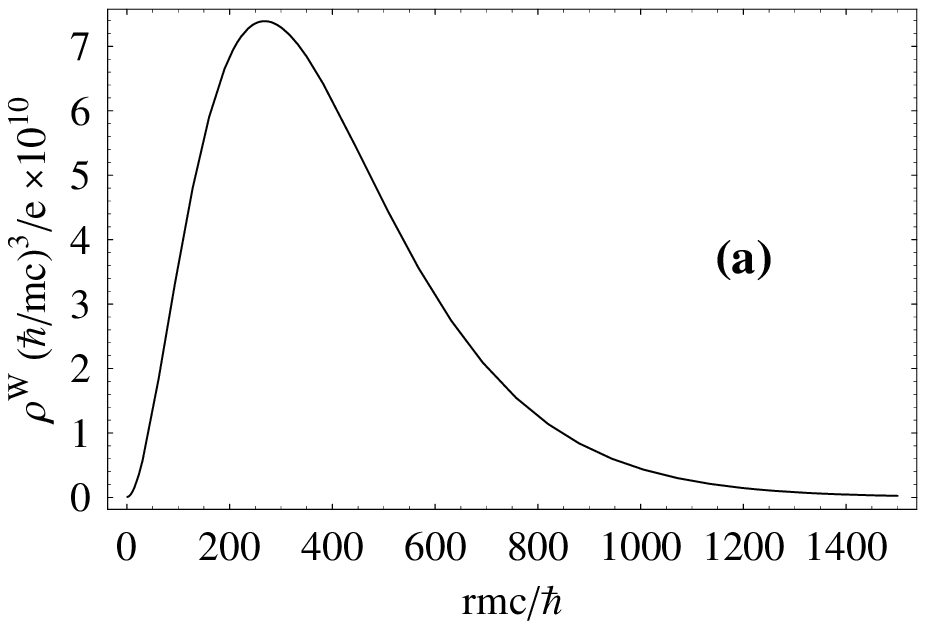}
\centering
 \includegraphics[height=5.3 cm,
keepaspectratio = true, 
]{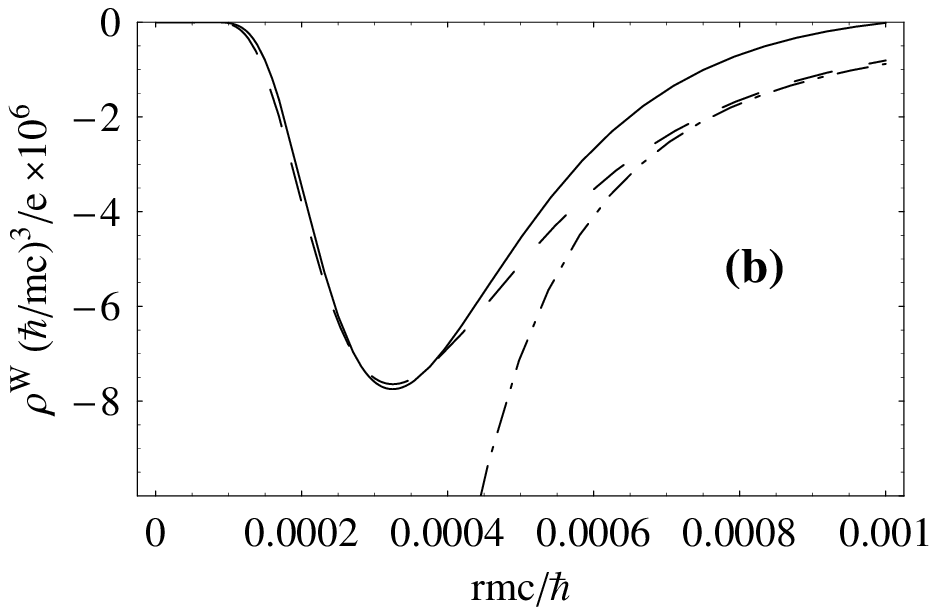}
\caption{ \label{two} The charge distribution $\rho^W(r)$ for the $2p_1$ state
  of a $W$-boson in the Coulomb field of a charge $Z=1$.  (a)
  Non-relativistic region of large distances $r\gg \hbar/mc$; (b)
  ultra-relativistic region $r\ll \hbar/mc$. Solid line - numerical
  solution, dashed line - analytical prediction of Eq.(\ref{rhoexp}),
  dashed-dotted line - the pure Coulomb case Eq.(\ref{cdvb}), when the
  charge density diverges at the origin (shown with arbitrary
  normalization).  }
\end{figure}
     \noindent 
     Using $S(r)$ from Eq.(\ref{Ueh}) one defines the potential $U(r)$
     Eq.(\ref{pot}) and the $\Upsilon$-potential in Eq.(\ref{Ups}).
     After that the solution of Eq.(\ref{radv}) is straightforward.
     This solution should be normalized on the total charge $e$ of the
     $W$-boson
      \begin{eqnarray}
        \label{norma}
        e=\int \rho^{W}(r)\, d^3r~, 
      \end{eqnarray}
      where the charge density is defined by Eq.(\ref{rode}). 
      
      Fig.\ref{one} shows the radial function $v(r)$ for the $2p_1$
      state ($n=2$, $j=0$) for $Z=1$.  For large distances $mr\gg 1$
      the function $v(r)$ is close to the conventional
      non-relativistic wave function of the $2p$ state, as it should
      be, see Section \ref{Nonrel}. In the ultra-relativistic region
      $mr \ll 1 $ the function $v(r)$ changes sign, then shows an
      extremum, and decreases exponentially when $r \rightarrow 0$, in
      accord with an analytical estimate Eq.(\ref{vexp}), which is
      shown by the dashed line.  The fact that the function $v(r)$ has
      a node in the relativistic region produces no controversy since
      $v(r)$ is not a proper wave function and the conventional
      theorem, which counts the nodes of the wave functions for
      discrete levels is not applicable.
     
      In our discussion we did not try to construct proper wave
      functions, being content with a possibility to calculate the
      current. As an example, Fig.\ref{two} shows the charge density
      for the $2p_1$ state in the Coulomb field of $Z=1$. In the
      non-relativistic region $mr\gg 1$ it behaves conventionally.
      For the ultra-relativistic case $mr\ll 1$ the density changes
      sign, exhibits an extremum and then decreases exponentially when
      $r\rightarrow 0$ in agreement with Eq.(\ref{rhoexp}).  Note that
      the ``wrong'' sign of the charge density, i. e.  the positive
      charge density for the negatively charged $W$-boson
      (Fig.\ref{two} shows $\rho^W/e=-\rho^W/|e|$) produces no
      contradiction with general principles. The Pauli's theorem, see
      e. g. Ref.  \cite{LL4}, implies that the energy of a boson field
      is positively defined, but the sign of the charge density of the
      boson field remains not unequivocally determined \cite{example}.
     
      The total positive charge located at small distances proves be
      very small; for $2p_1$, $Z=1$ state it is
     \begin{eqnarray}
       \label{wrong}
       Q^W_\mathrm{+}=4\pi \int_0^{r_0} \rho^W(r)\,r^2\,dr\simeq
       0.676\cdot 10^{-14}\,|e|~.
     \end{eqnarray}
     Here $r_0\simeq 1.01\cdot 10^{-3}m^{-1}$ is a node of $\rho^W
     (r)$.
     
     It is instructive to compare the found charge density with the
     one in the pure Coulomb problem, which is shown in Fig.
     \ref{two} (b) by the dashed-dotted line that reproduces
     Eq.(\ref{cdvb}). One should be content in this case with an
     arbitrary normalization, since for the pure Coulomb field the
     normalization integral in Eq.(\ref{norma}) is divergent. Fig.
     \ref{two} (b) illustrates the fact that the vacuum polarization reduces
     the charge density at the origin.
      
     The energy shift $\delta \varepsilon$ of the level $2p_1$
     ($\delta \varepsilon$ is a deviation of energy from the
     Sommerfeld formula Eq.(\ref{Mspe})) due to the vacuum
     polarization is found to be $\delta \varepsilon/m=-1.90 \cdot
     10^{-7}$. In relative units it is much bigger than the Lamb shift
     in atoms $\delta \varepsilon_\mathrm{LS}/m_e\sim Z^4 \alpha^5$.
     The reason is obvious. The Lamb shift in atoms originates mostly
     from within the Compton distances $r\sim r_c= 1/m_e$, which are
     smaller than the Bohr radius for the electron $r_\mathrm{B}\sim
     1/(Z\alpha m_e)$.  For $W$-bosons the situation is different.
     Light fermions, which contribute to Eq.(\ref{Ueh}), allow the
     polarization potential to spread to large distances, as far as
     the Bohr radius of the $W$-boson, $r \sim 1/m_i \sim 1/(Z\alpha
     m)$. Therefore the energy shift due to the vacuum polarization
     gains substantial contribution from the non-relativistic region,
     where the wave function is large, which makes $\delta
     \varepsilon$ large as well (large compared to the Lamb shift in
     atoms).  The accuracy of the energy shift calculations is limited
     by an accuracy of our model at large distances.  The contribution
     of the QCD vacuum, which could be substantial here, is not
     described properly by the model based on Eq.(\ref{Ueh}).
     Consequently, the presented above value for the energy shift
     should be considered only as an estimate.
     
     Nevertheless, one can derive an important lesson from this
     estimate.  The found energy shift is small on the absolute scale,
     being lower than the non-relativistic binding energy by a factor
     of $|\delta \varepsilon|\times 4/(Z^2\alpha^2m) \sim 1.4\cdot
     10^{-2}$.  Thus, the dramatic variation of the function $v(r)$ at
     the origin, which is produced by the vacuum polarization, makes
     only small impact on the spectrum.  This is in contrast to a
     strong influence, which the vacuum polarization exercises on the
     charge distribution of vector bosons.  The fact that the energy
     shift is small makes the Sommerfeld formula Eq.(\ref{Mspe}) a
     good approximation for discrete energy levels.

      \section{Discussion}
      \label{conclusion}
      
      We demonstrated that the conventional QED vacuum polarization
      plays a very important, defining role in the Coulomb problem for
      vector bosons. Let us summarize the reasons leading to this
      conclusion.  The Uehling potential, which describes the vacuum
      polarization in the simplest approximation is known to be a
      weakly attractive and slowly varying function. For spinor
      particles it produces a small enhancement of the fermion wave
      functions on the Coulomb center. For vector bosons the situation
      is different because the equations of motion for vector
      particles explicitly incorporate the external current. As a
      result, the density of the polarized charge $\rho$ comes into
      the equations of motion for vector bosons. The corresponding
      term in the equations was called the $\Upsilon$-potential,
      $\Upsilon =e\rho/m^2$. The charge density $\rho$ is negative for
      an attractive Coulomb center, $\rho<0$ when $Z>0$, being
      singular on the Coulomb center, $|\rho| \sim 1/r^3$. One derives
      from this that the vacuum polarization produces a repulsive
      $\Upsilon$-potential, $\Upsilon=e\rho/m^2 \propto 1/r^3 >0$
      (remember, $e<0$). Since the $\Upsilon$-term is singular at the
      origin, it plays a dominant role at small distances.
      
      Strong effective repulsion produced by the $\Upsilon$-potential
      reduces the fields, which describe $W$-bosons on the Coulomb
      center. For $j=0$ this reduction is dramatic, exponential. For
      $j=1$, ''$\gamma-3/2$'' the suppression is of a more moderate
      power-type nature, but in both cases it is strong enough to
      eliminate the infinite charge, which is located at the origin in
      the pure Coulomb approximation.
      
      The above comments appeal to a chain of calculations.  It is
      interesting to look at the obtained result from a more general
      perspective.  The renormalizabity of the Standard Model implies
      that by renormalizing relevant physical quantities one is {\em
        bound} to obtain sensible physical results. The relevant
      quantity in question is the charge density of a vector boson.
      It follows from this that the important physical quantity, which
      should be renormalized, is the coupling constant.  Its
      renormalization is effectively fulfilled when the vacuum
      polarization is taken into account. Thus, it makes sense that
      the account of the vacuum polarization results in acceptable
      physical results. 
      
      A proposed approach is very straightforward, which makes the
      Coulomb problem for vector bosons as simple and reliable as it
      is for scalars and spinors.  All discrete energy levels can be
      easily evaluated, all relevant fields can be calculated and
      normalized properly.  Presumably, all scattering data can also
      be evaluated, though the scattering problem was not discussed in
      detail in the present work. All these quantities include the
      Coulomb charge $Z$ accurately, not relying on perturbation
      theory.  Starting from this base, one can consider all other
      processes left outside the scope of the Coulomb problem by
      treating them as perturbations. This includes the conventional
      QED processes, such as the radiative decay, photoionization, the
      radiative corrections.  This includes also processes related to
      possible exchange of Higgs and $Z$-bosons.
            
      Previous attempts to formulate the Coulomb problem for vector
      bosons within the framework of the Standard Model have been
      facing a difficulty related to an infinite charge of the boson
      located near an attractive Coulomb center.  This work finds that
      the polarization of the QED vacuum eradicates the problem.
      Usually the QED radiative corrections produce only small
      perturbations. It is interesting that in the case discussed the
      radiative correction plays a major, defining role.

      
      This work was supported by the Australian Research Council. One
      of us (VF) appreciates support from Department of Energy, Office
      of Nuclear Physics, Contract No.  W-31-109-ENG-38.

    \appendix

    \section{Homogeneous magnetic field}
    \label{homo}
    For a static homogeneous magnetic field Eq.(\ref{form}) reads
    \begin{eqnarray}
      \label{mag}
      \left( \varepsilon^2-m^2\right)\boldsymbol{W} =
      -\left( \boldsymbol{\nabla}-ie \boldsymbol{ A}\right)^2\boldsymbol{ W}-2ie
      \boldsymbol{B}\times \boldsymbol{W}.~~
    \end{eqnarray}
    Assuming that the magnetic field is directed along the $z$-axis
    and introducing the new variables $w_\sigma,~\sigma=0,\pm1$, $
    w_{\pm 1}=(\boldsymbol{ W}_x\pm i\boldsymbol{ W}_y)/\sqrt 2$, $
    w_{0}=\boldsymbol{ W}_z$ one rewrites Eq.(\ref{mag}) in a simple
    form
    \begin{eqnarray}
      \label{wsig}
      \left( \varepsilon^2-m^2\right)\,w_\sigma =
      -\left( \boldsymbol{ \nabla}-ie \boldsymbol{ A}\right)^2w_\sigma
      +
      2eB\sigma\,w_\sigma~,
    \end{eqnarray}
    which looks similar to the non-relativistic Schr\"odinger equation
    for a particle in the homogeneous magnetic field. This similarity
    allows one to write the spectrum Eq.(\ref{e2}).

 \section{relativistic corrections to energy levels}
 \label{appendix}

 Here we present separate expectation values for four relativistic
 corrections in the same order as they appear in Eq.(\ref{dH}). For
 $l=0$
  \begin{eqnarray}
      \label{nl=0j}
      \delta E_{n0j}= \frac{m(Z\alpha)^4}{n^3}\left[
        \left(\frac{3}{8n}-1\right)+0+ \frac{2}{3}+0\right]~.
    \end{eqnarray}
    Here and below we specify the terms having zero expectation values
    by writing the corresponding zeros explicitly.  For $l\ge 1 $
  \begin{eqnarray}
\nonumber
      \delta E_{nlj}= \frac{m(Z\alpha)^4}{n^3} \left[\,
        \left(\frac{3}{8n}-\frac{1}{2l+1} \right)\right. + 
\\       \label{nl>0j}
 \frac{\langle ls\rangle }{l(l+1)(2l+1)}
 +0 \\ \nonumber
\left.
 - \frac{6 \langle ls \rangle^2+3\langle ls 
 \rangle-4l(l+1)}{l(l+1)(2l-1)(2l+1)(2l+3)}\,\right]~,
    \end{eqnarray}
    where
  \begin{equation}
      \label{ls}
      \langle ls \rangle=\frac{1}{2}\,[j(j+1)-l(l+1)-2]~.
    \end{equation}
    Both Eq.(\ref{nl=0j}) and Eq.(\ref{nl>0j}) lead to Eq.(\ref{nlj}).
    The total relativistic correction would look very complicated and
    show no degeneracy if magnetic dipole or electric quadrupole
    moments of a vector particle are different from those values,
    which follow from the gauge theory.

    \section{Spherical vectors}
    \label{spherical}
    The conventional definition of spherical vectors, see e.g.
    \cite{LL4}, reads
    \begin{eqnarray}
      \label{Y}
      \begin{array}{l}
      \boldsymbol{ Y}^{(e)}_{jm}= \boldsymbol{
        \nabla}_n\,Y_{jm}/\sqrt{j(j+1)}~,
      \\
      \boldsymbol{ Y}^{(l)}_{jm}= \boldsymbol{ n} \,Y_{jm}~,
      \\ 
      \boldsymbol{ Y}^{(m)}_{jm}= \boldsymbol{ n \times Y}^{(e)}_{jm}
        ~.
      \end{array}
    \end{eqnarray}
    Here $Y_{jm}\equiv Y_{jm}(\theta,\varphi)$ is the spherical
    function, $\boldsymbol{ Y}^{(e)}_{jm},\boldsymbol{ Y}^{(l)}_{jm},
    \boldsymbol{ Y}^{(m)}_{jm}$ are the electric, longitudinal and
    magnetic vectors. The symbol $\boldsymbol{ \nabla}_n$ in
    Eq.(\ref{Y}) indicates the angular part of the gradient, $
    \boldsymbol{ \nabla}F(\theta,\phi)= \boldsymbol{ \nabla}_n
    F(\theta,\phi)/r$, and $\boldsymbol{ n}=\boldsymbol{ r}/r$ is a
    unit vector along the radius vector.

    Definitions Eqs.(\ref{Y}) imply the following properties of the
    spherical vectors
    \begin{eqnarray}
      \label{div}
         \begin{array}{l}
        \boldsymbol{ \nabla }_n \cdot \boldsymbol{ Y}^{(e)}_{jm}
        =-\sqrt{j(j+1)}\,\,Y_{jm}\,,
        \\
        \boldsymbol{ \nabla}_n \cdot \boldsymbol{ Y}^{(l)}_{\!jm}
        =2\,Y_{jm}\, ,
        \\
        \boldsymbol{ \nabla}_n \cdot \boldsymbol{ Y}^{(m)}_{jm}=0\,. 
      \end{array} 
      \\ \label{rot}
        \begin{array}{l}
       \boldsymbol{ \nabla \times Y}^{(e)}_{jm}=
        \boldsymbol{ Y}^{(m)}_{jm}\, ,
      \\ 
      \boldsymbol{ \nabla \times Y}^{(l)}_{jm}=
        -\sqrt{j(j+1)}\,\boldsymbol{ Y}^{(m)}_{jm}~,
        \\       
      \boldsymbol{ \nabla \times Y}^{(m)}_{jm}=
        -\boldsymbol{ Y}^{(e)}_{jm}-\sqrt{j(j+1)}\,\boldsymbol{ Y}^{(l)}_{jm}~.
  \end{array} 
    \end{eqnarray}
    The formulas for the Laplace operator read
      \begin{eqnarray}
        \label{ddd}        
     \begin{array}{l}
     \Delta_n\boldsymbol{ Y}^{(\,e\,)}_{jm}  = -j(j+1)\,\boldsymbol{ Y}^{(e)}_{jm}+
          2\sqrt{j(j+1)} \, \,\boldsymbol{ Y}^{(l)}_{jm},
          \\ 
           \Delta_n\boldsymbol{ Y}^{(\,l\,)}_{jm}=2\sqrt{j(j+1)}\,\,
          \boldsymbol{ Y}^{(e)}_{jm}
          - \big(j(j+1)+2\big) \boldsymbol{ Y}^{(l)}_{jm},
          \\ 
          \Delta_n\boldsymbol{ Y}^{(m)}_{jm} =-j(j+1)\boldsymbol{
          Y}^{(m)}_{jm}.
        \end{array}
    \end{eqnarray}
    Here $\Delta_n$ describes the angular part of the Laplacian, i.e.
    $\Delta F(\theta,\phi) =\Delta_nF/r^2$. The parity for electric
    and longitudinal polarizations equals $P=(-1)^{j}$, for magnetic
    polarization the parity is $P=(-1)^{j+1}$. The orbital moment $l$
    takes the value $l=j$ for the magnetic polarization, in agreement
    with the parity for this state. The electric and longitudinal
    polarizations are constructed as linear combinations of the two
    states with $l=j\pm 1$. For $j=0$ there exists only one spherical
    vector, which is purely longitudinal and has $l=1$.

    \section{Spectrum  of electro-longitudinal modes for $j\ge 1$}
    \label{non0}
    
    Let us verify that for $j\ge 1$ the function $g$ introduced in
    Eq.(\ref{chi}) is nonzero.  Consider first the ``$\gamma-1/2$''
    mode. Substituting Eq.(\ref{uas}) into Eq.(\ref{chi}) one finds
        \begin{eqnarray}
      \label{chiA}
      g\rightarrow a\frac{1}{Z\alpha}(1/4-\gamma^2)\,r^{\gamma-1/2}~,
    \end{eqnarray}
    which indicates that in this mode $g$ is not zero.
    
    Consider now the ''$\gamma-3/2$'' mode, which incorporates both
    possible polarizations at small distances.  We need here the
    expressions for $u$ and $v$ at small distances that are more
    accurate, then the ones in Eq.(\ref{nu}). They can be derived by
    using $mr\ll 1$ as a perturbation in
    Eqs.(\ref{sim1}),(\ref{sim2}), and pushing calculations one step
    beyond the simplest approximation given by
    Eqs.(\ref{nu}),(\ref{banu}). The result reads
    \begin{eqnarray}
      \label{anext}
      && u\rightarrow  b\left( 1-\frac{2+(Za)^2}{\gamma+1/2}
        \,\,\frac{\varepsilon r}{Z\alpha}\right)r^{\gamma-3/2}~,\\
      \label{bnext}
      && v\rightarrow \frac{      b  }{ \sqrt{j(j+1)} }\left(
        \gamma-\frac{1}{2}
        -Z\alpha \varepsilon r
      \right)r^{\gamma-3/2}.~
    \end{eqnarray}
    Substituting Eqs.(\ref{anext}),(\ref{bnext}) into Eq.(\ref{chi})
    one finds that the main term $\propto r^{\gamma-3/2} $ cancels out
    in $g$, but the next one survives, giving
    \begin{eqnarray}
      \label{chiB}
      g \rightarrow b \frac{\varepsilon}{Z\alpha}
      \left(2\gamma-1-(Z\alpha)^2\right)\,r^{\gamma-1/2}~.
    \end{eqnarray}
    We verified that for $ j\ge1$ the function $g$ is not an identical
    zero for both electro-longitudinal modes. 

    \section{Longitudinal mode $j=0$ and matrix mechanics}
    \label{matrix}
    
    In order to find the spectrum of the operator $H$ in Eq.(\ref{H})
    let us employ a method, which finds its inspiration in an elegant
    approach to quantum mechanics developed by the G\"otingen School
    and often called the {\it matrix mechanics}; the book Ref.
    \cite{green} gives its systematic presentation. We modify it for
    our purposes as follows.  Assume that one needs to find discrete
    spectrum of some Hermitian operator ${\mathcal H}$ (in our case it is the
    operator ${\mathcal H}$ in Eq.(\ref{H})).  Let us presume that one is able to
    find the operator $\theta$, which satisfies
    \begin{eqnarray}
      \label{theta}
      {\mathcal H}=\theta^\dag\theta+\lambda_0~,
    \end{eqnarray}
    where $\lambda_0$ is a number. Define then a new operator $\tilde {\mathcal H}$,
    \begin{eqnarray}
      \label{m+1}
      \tilde {\mathcal H}=\theta\,\theta^\dag+\lambda_0~.
    \end{eqnarray}
    Let us verify that the two operators ${\mathcal H},\tilde {\mathcal H}$ have very similar sets
    of eigenvalues.  Consider an eigenfunction $\varphi$ of ${\mathcal H}$, with
    the eigenvalue $\lambda$
    \begin{eqnarray}
      \label{psin}
      {\mathcal H}\varphi=\lambda\,\varphi~.
    \end{eqnarray}
    Taking 
    \begin{eqnarray}
      \label{psi2old}
      \tilde \varphi=\theta \varphi~, 
    \end{eqnarray}
    one verifies that
    \begin{eqnarray}
      \label{psi1old}
      \tilde {\mathcal H}\,\tilde \varphi=(\theta\,\theta^\dag+ \lambda_0)\theta\,\varphi=
      \theta\,(\theta^\dag\,\theta+\lambda_0)\varphi
      \\ \nonumber
      =\theta\,{\mathcal H}\varphi=\lambda\,\theta\,\varphi=\lambda\tilde \varphi~.
    \end{eqnarray}
    This shows that either the function $\tilde \varphi$ is an eigenfunction of
    $\tilde {\mathcal H}$ with the eigenvalue $\lambda$, or $\tilde \varphi=0$.  The first
    options makes $\lambda$ an eigenvalue of both operators ${\mathcal H},\tilde {\mathcal H}$.
    The second one implies that
    \begin{eqnarray}
      \label{impl}
      \theta\,\varphi=0~,
    \end{eqnarray}
    which indicates that $\lambda =\lambda_0$ is a candidate for an
    eigenvalue of ${\mathcal H}$ because Eq.(\ref{impl}) implies
    ${\mathcal H}\,\varphi=\lambda_0 \varphi$. Eq.(\ref{impl}) provides a
    convenient way to derive the corresponding eigenfunction. There is
    though a subtlety here. The found from Eq.(\ref{impl}) $\varphi$
    may, or may not satisfy the boundary conditions. If it does, then
    it represents the eigenfunction and $\lambda=\lambda_0$ is an
    eigenvalue. Otherwise, $\lambda_0$ does not belong to the discrete
    spectrum, as would be the case in an example discussed.  One
    should also verify that an action of the operator $\theta$ in
    Eq.(\ref{psi2old}) (as well as the operator $\theta^\dag$ in
    Eq.(\ref{then}) below) does not spoil the boundary conditions.  We
    presume here that this is the case, and verify later on that this
    assumption holds for a particular example discussed, see
    Eq.(\ref{sens1}),(\ref{sens2}).
    
    We conclude that any discrete eigenvalue of ${\mathcal H}$ is an eigenvalue
    of $\tilde {\mathcal H}$ as well, with one possible exception of $\lambda_0$.  By
    reversing the argument, one derives that if $\tilde \varphi$ is an
    eigenfunction of $\tilde {\mathcal H}$ with the eigenvalue $\tilde\lambda$,
    \begin{eqnarray}
      \label{l'}
      \tilde {\mathcal H}\tilde \varphi=\tilde\lambda\tilde \varphi
    \end{eqnarray}
    then
    \begin{eqnarray}
      \label{then}
      \varphi=\theta^\dag\tilde \varphi
    \end{eqnarray}
    satisfies Eq.(\ref{psin}) with $\lambda=\tilde\lambda$.  We see
    that the two sets of discrete eigenvalues of the two operators $
    {\mathcal H},\tilde {\mathcal H} $ are same, except for $\lambda_0$, which may, or may
    not be present in one, or both sets of spectra. The crucial for us
    point is that the operator $\tilde {\mathcal H}$ can be more simple for
    analyses than the initial operator ${\mathcal H}$.
 
    Taking the operator ${\mathcal H}$ from Eq.(\ref{H}), we construct the
    operators $\theta,\theta^\dag$ in the form
    \begin{eqnarray}
      \label{the}
      \theta&=&-\frac{d}{dx}+a+\frac{b}{x}+\frac{c}{x+1}~,
      \\ \label{the'}
      \theta^\dag &=&~~\,\frac{d}{dx}+a+\frac{b}{x}+\frac{c}{x+1}~,
    \end{eqnarray}
    where $a,b,c$ are real numbers. From Eqs.(\ref{the}),(\ref{the'})
    it follows that
    \begin{eqnarray}
      \label{thth}
      \theta^\dag\theta=-\frac{d^2}{dx^2}+a^2&+&\frac{ 2b(a+c) }{ x }
      +\frac{ b(b-1) }{x^2}\\ \nonumber
      &&+\frac{ 2c(a-b)}{x+1}  +\frac{c(c-1)}{(x+1)^2}~.
    \end{eqnarray}
    There are four $x$-dependent rational terms in Eq.(\ref{thth}),
    while only three coefficients $a,b,c$ are available for tuning to
    make them identical to similar terms present in the operator ${\mathcal H}$.
    However, the coefficients in Eq.(\ref{H}) prove to be
    ``user-friendly'', making this procedure possible. Taking
    \begin{eqnarray}
      \label{abc}
      &a=b=\gamma+1/2\,,\quad \quad c=-1~,
      \\ \label{lambda0}
      &\lambda_0=-\left(\gamma+1/2\right)^2~,
    \end{eqnarray}
    one satisfies Eq.(\ref{theta}).  Taking $\theta,\theta^\dag$
    defined in Eqs.(\ref{the}),(\ref{the'}) and (\ref{abc}) one
    constructs $\tilde {\mathcal H}$ Eq.(\ref{psi1old}), with the result given in
    Eq.(\ref{psi1}). The ``nasty'' singular at $x=-1$ term disappears
    from $\tilde {\mathcal H}$. The latter operator describes a conventional
    Coulomb-type problem with $L=\gamma+1/2 $ playing a role of an
    effective (non-integer) angular momentum.  From Eq.(\ref{H2}) one
    finds that regular at $x=0$ solution of the eigenvalue problem
    $\tilde {\mathcal H}\tilde \varphi=-\varkappa^2\,\tilde \varphi$, satisfies
    Eq.(\ref{psi2in}).  Eq.(\ref{Ry}), which ensures that this
    solution is regular at infinity, completely defines a set of
    discrete eigenvalues of $\tilde {\mathcal H}$.
    
    The set of eigenvalues of $\tilde {\mathcal H}$ gives the eigenvalues of the
    original operator ${\mathcal H}$, except for possibly one additional
    eigenvalue $\lambda_0$, which is discussed below.  The
    eigenfunctions of ${\mathcal H}$ can be found from Eq.(\ref{then}).  Using
    Eqs.(\ref{the'}),(\ref{abc}) one presents them in a form of
    Eq.(\ref{for}).
    
    In order to verify whether $\lambda_0$ is an eigenvalue of ${\mathcal H}$ one
    needs to find $\varphi$ from Eq.(\ref{impl}). Eq.(\ref{the}) gives
    \begin{eqnarray}
      \label{l0}
      \left(-\frac{d}{dx} +\left 
          (\gamma+1/2\right)\frac{x+1}{x}- \frac{1}{x+1}\right)\varphi=0~,
    \end{eqnarray}
    which leads to
    \begin{eqnarray}
      \label{New-zero}
      \varphi=(x+1)^{-1}\,x^{\gamma+1/2}\,\exp\,[\,(\gamma+1/2)\,x\,]~.
    \end{eqnarray}
    Since this function is singular at $x=\infty$, it cannot be an
    eigenfunction. Consequently $\lambda_0$ is not an eigenvalue.
    
    The function $\varphi$ defined by Eq.(\ref{for}) exhibits regular
    behavior at both boundaries Eqs.(\ref{sens1}),(\ref{sens2}).  This
    ensures that $\varphi$ is an eigenfunction.  Note, that specifying
    the operators $\theta,\theta^\dag$ one had an additional option.
    One could have chosen in Eqs.(\ref{abc}) and all the following
    relevant formulas $-\gamma$ instead of $\gamma$.  It this case,
    however, instead of Eq.(\ref{sens2}) one obtains $\varphi \propto
    x^{-\gamma+1/2}$ for $x\rightarrow 0$, which indicates a singular,
    unacceptable for an eigenfunction behavior.
    
    We conclude that the full set of all discrete eigenvalues of ${\mathcal H}$
    is specified by Eq.(\ref{Ry}). The corresponding eigenfunctions
    are given by Eqs.(\ref{for}),(\ref{psi2}).


\begin{thebibliography}{99}
\bibitem{example}
    
      The scalar field $\phi(r)$ propagating in a static potential
      $U(r)$ provides a simple example. The charge density
      $\rho_\phi=2e(\epsilon-U(r))\phi^*(r)\phi(r)$, where $\epsilon$
      is the energy, shows the ``wrong'' sign in a strongly repulsive
      region $\epsilon<U(r)$. It is necessarily small because the
      field is suppressed their; compare Eq.(\ref{wrong}), which
      indicates that the ``wrong'' charge of the $W$-boson in the
      $2p_1$ state is very small.
\end{thebibliography}
\end{document}